\documentclass[]{aa}  

\usepackage[varg]{txfonts}
\usepackage{placeins}
\usepackage{natbib}
\usepackage{url}
\PassOptionsToPackage{hyphens}{url}
\usepackage{graphicx}
\usepackage{subfigure}
\usepackage{comment}
\usepackage{amsmath}
\usepackage{txfonts}
\usepackage{lipsum}
\usepackage{xcolor,soul}
\sethlcolor{lightgray}
\usepackage{listings} 
\usepackage{times}
\usepackage{multirow}
\usepackage{booktabs}
\usepackage{float}
\usepackage{epstopdf}
\usepackage{longtable}
\usepackage{soul}
\usepackage[colorlinks=true,linkcolor=red,citecolor=blue, breaklinks=true]{hyperref}
\restylefloat{figure}  
\usepackage{lscape}

\usepackage{linenoaa}
\usepackage{lineno}

\newcommand{\herschel}{{\it Herschel}}

\def\NHUNIT{\ifmmode {\rm \,cm^{-2}} \else $\rm \,cm^{-2}$ \fi} 

\def\nhh{\ifmmode N_{\rm H_{2}}\else $N_{\rm H_{2}}$\fi} 
\def\nhhc{\ifmmode N_{\rm H_{2}}^0\else $N_{\rm H_{2}}^0$\fi} 
\def\nhhbg{\ifmmode N_{\rm H_{2}}^{\rm bg}\else $N_{\rm H_{2}}^{\rm bg}$\fi} 
\def\nh{\ifmmode N_{\rm H}\else $N_{\rm H}$\fi}
\def\ml{\ifmmode M_{\rm line}\else $M_{\rm line}$\fi}  
\def\sunpc{\ifmmode \rm \,M_\odot/\rm pc\,\else $\rm \,M_\odot/\rm pc\,$\fi}  

\def\rflat{\ifmmode R_{\rm flat}\else $R_{\rm flat}$\fi}   
\def\kms{\ifmmode {km\,s$^{-1}$}\else km\,s$^{-1}$\fi}  
\def\n2dp{\ifmmode {N$_2$D$^+$}\else {N$_2$D$^+$}\fi}  
\def\pnh2d{\ifmmode {pNH$_2$D}\else {pNH$_2$D}\fi}  

\def\arcm{\ifmmode {^{\scriptstyle\prime}}
          \else $^{\scriptstyle\prime}$\fi}
\def\arcs{\ifmmode {^{\scriptstyle\prime\prime}}
          \else $^{\scriptstyle\prime\prime}$\fi}
\newdimen\sa  \newdimen\sb
\def\parcs{\sa=.07em \sb=.03em
     \ifmmode \hbox{\rlap{.}}^{\scriptstyle\prime\kern -\sb\prime}\hbox{\kern -\sa}
     \else \rlap{.}$^{\scriptstyle\prime\kern -\sb\prime}$\kern -\sa\fi}
\def\parcs{\sa=.07em \sb=.03em
     \ifmmode \hbox{\rlap{.}}^{\scriptstyle\prime\kern -\sb\prime}\hbox{\kern -\sa}
     \else \rlap{.}$^{\scriptstyle\prime\kern -\sb\prime}$\kern -\sa\fi}
\def\parcm{\sa=.08em \sb=.03em
     \ifmmode \hbox{\rlap{.}\kern\sa}^{\scriptstyle\prime}\hbox{\kern-\sb}
     \else \rlap{.}\kern\sa$^{\scriptstyle\prime}$\kern-\sb\fi}
\def\parcd{\sa=.08em \sb=.03em
     \ifmmode \hbox{\rlap{.}\kern\sa}^{\scriptstyle\circ}\hbox{\kern-\sb}
     \else \rlap{.}\kern\sa$^{\scriptstyle\circ}$\kern-\sb\fi}
\newcommand{\rev}[1]{{#1}}
\newcommand{\revn}[1]{{#1}}
\newcommand{\revl}[1]{{#1}}

\begin{document} 

\title{Probing the ion-neutral drift velocity towards the L1544 prestellar core\thanks{This work is based on observations carried out with the IRAM 30m telescope. IRAM is supported by INSU/CNRS (France), MPG (Germany) and IGN (Spain).}}
\subtitle{Detection of ambipolar diffusion using \n2dp\ and para-NH$_2$D}

 \titlerunning{Probing the ion-neutral drift velocity towards the L1544 prestellar core}

\author{Doris Arzoumanian\inst{1,2,3}\fnmsep\thanks{Corresponding author: arzoumanian.doris.958@m.kyushu-u.ac.jp}
  \and
Silvia Spezzano\inst{4}
  \and
  Tommaso Grassi\inst{4}
  \and
  Paola Caselli\inst{4}
  \and
  Yusuke Tsukamoto\inst{5}    
  \and
  Haruka Fukihara\inst{5}   
    \and
  Yoshiaki Misugi\inst{6,3}    
    \and
  Felipe Alves\inst{7}    
    \and
  Jaime Pineda\inst{4}
      \and
  Sigurd Jensen\inst{4}
    \and
  Elena Redaelli\inst{8}
      \and
  Alexei Ivlev\inst{4}
  }
      
   \institute{Institute for Advanced Study, Kyushu University, Japan
     \and
Department of Earth and Planetary Sciences, Faculty of Science, Kyushu University, Nishi-ku, Fukuoka 819-0395, Japan
  \and
National Astronomical Observatory of Japan, Osawa 2-21-1, Mitaka, Tokyo 181-8588, Japan
  \and
Max-Planck-Institut f\"ur Extraterrestrische Physik, Giessenbachstrasse 1, 85748 Garching, Germany
  \and
Kaghoshima University, Department of Physics and Astronomy Graduate School of Science and Engineering, Korimoto, Kagoshima, Japan
  \and
Faculty of Science and Engineering, Kyushu Sangyo University, 2-3-1 Matsukadai, Fukuoka 813-8503, Japan
  \and
Institut de Radioastronomie Millim\'etrique (IRAM), 300 rue de la Piscine, F-38406, Saint-Martin d’H\`eres, France
  \and
European Southern Observatory, Karl-Schwarzschild-Stra{\ss}e 2, 85748 Garching, Germany}

     \date{}

\abstract
{The dynamical role  of the magnetic field in the star formation process is tightly linked to the coupling between matter and the field. This coupling is due to the interaction between ions and neutrals in the partially ionized interstellar medium. When the ionization degree drops in the dense environment of prestellar cores, the magnetic field and the matter may decouple, leading to differences in the infalling velocities  of ions and neutrals known as ambipolar diffusion. }%
 {The onset of gravitational collapse  resulting from  ion-neutral decoupling has never been observed. 
 The aim of this work is to search for signatures {\revn of  ambipolar diffusion within a prestellar core.} 
 }
 %
   {We {\revn observed} the deuterated  \n2dp\ ion and the neutral para-NH$_2$D species towards the prototypical prestellar core L1544. 
   These two species are ideal tracers of prestellar cores sampling the same high densities in the core interior.
{\revn We compared the  velocity centroid and linewidth maps of the ion-neutral pair.}
}
{  
We find a mean ion-neutral velocity difference  of $\sim0.05$\,km/s towards the core. 
By comparing with predictions from  self-consistent calculations of the ambipolar resistivity including dust grain growth, %
    we interpret the observed  ion-neutral velocity difference in L1544 as a signature of ambipolar diffusion.
     {\revn We do not detect a significant ion-neutral linewidth difference that may be attributed to the subsonic infall motions of the gas in L1544 and geometrical effects in the presence of inclination.}
 }
  {  
  These results emphasize the role of dust grain growth at the  prestellar core stage in setting the ambipolar resistivity and regulating the dynamical evolution of dense cores towards their collapse into protostars. We propose that measurements of ion-neutral drift velocities   provide new constraints on the total magnetic field strength and the  dust size distribution within prestellar cores.}


\maketitle
\nolinenumbers

\section{Introduction}\label{intro}

Prestellar cores are gravitationally bound objects with a centrally concentrated density distribution (peak densities $>10^5$\,cm$^{-3}$) on the verge of star formation, but they do not host any protostar at their center \citep{Ward-Thompson1999,Keto2008,Pineda2023}. The gravitational collapse of  prestellar cores will result in stellar systems. 
Prestellar cores provide the pristine  conditions before the formation of a protostellar seed and the launching of the outflow altering the chemical and physical conditions at the onset of collapse. 
Consequently, the study of prestellar core properties is extremely important in revealing the initial physical and chemical conditions needed to understand how stellar systems like our own are assembled.
%
Protostars form from the gravitational collapse of prestellar cores when gravity overcomes {\rev opposing 
forces, i.e., 
thermal pressure, turbulence, and magnetic pressure and tension.}
Observations of core velocity dispersions often show narrow linewidth, which are on the order of the thermal velocity dispersions ($\sim 0.2\,$km/s for a molecular gas at 10\,K), suggesting sub/transcritical turbulence in prestellar cores \citep{Goodman1998,Pineda2010,Pineda2025}. 

Although inferring the magnetic field (B-field) geometry and strength from observations is challenging, in the past five to ten years significant progress has been made to describe the B-field structure in molecular clouds, filaments, and dense cores \citep[cf. recent reviews by][]{Hull2019,Maury2022,Pattle2023}. 
{\rev While observationally driven estimates of the mass-to-flux ratio may be hindered by limitations \citep[e.g.,][]{Tritsis2025}, 
estimates of magnetic energies derived from} polarization observations 
towards filaments and cores, are often comparable to the {\rev gravitational energies \citep[e.g.,][]{Pattle2023}, suggesting that B-fields may play an important role in regulating core collapse and star formation \citep[see also the simulations by][]{Nakamura2008}. 
 Recent studies comparing  molecular line observations and synthetic maps from magnetohydrodynamic (MHD) simulations  also suggest that  protostars may  form  from the collapse of magnetically sub/trans-critical cores \citep[cf.][]{Yin2021,Priestley2022}.}

To allow for gravitational collapse of the cores, the B-field must diffuse.
The decoupling between neutral and charged particles has been proposed to be an efficient  {\rev diffusion} mechanism of the B-field during the prestellar core collapse and the early phases of  protostellar evolution. This process, known as ambipolar {\rev diffusion  \citep{Mestel1956,Mouschovias1979,Mouschovias1987b} causes} the neutral particles to drift through magnetic field lines allowing the matter to be (quasi-statically) accreted into the core \citep{Mouschovias2011}. 

In the classical model of a collapsing magnetized core, an  {\it hourglass} B-field morphology pinched towards the centre of mass is expected. 
This pinched hourglass shape is formed by the material collapsing freely along the field lines, and the flux-frozen field being dragged in by the collapse across the lines \citep{Mestel1966,Myers2018}.  
The morphology of the B-field lines is related to the importance of coupling between matter and flux, where weaker flux-freezing leads to
stronger ambipolar diffusion  and hence less field curvature, i.e., less pinched B-field lines \citep[e.g.,][]{Basu2009,Myers2018}.
Observations show a few cases of hourglass magnetic field structures towards  
{\rev massive dense cores \citep{Girart2009,Beltran2019,Saha2024} and  low mass cores \citep{Girart2006,Frau2011,Alves2011,Stephens2013,Redaelli2019Bfield}.} However, in a large number of collapsing cores  these pinched magnetic field structures are not observed.
While projection effects and line-of-sight integration may partly  hinder the detection of these hourglass B-field shapes, their non-detection may indicate   a partial decoupling of the matter and the field at the core scale allowing  the matter to infall towards the center through ambipolar diffusion without dragging in the B-field lines and consequently increasing the B-field flux.

 In spite of its importance, detecting 
 the ion-neutral drift in observations 
 is a challenging task: 1) the small predicted drift velocity (a fraction of the sound speed) requiring high-spectral resolution data \citep{Tritsis2022}. 2) The observed charged and neutral molecular species need to be chemically co-evolving to be tracing the same gas, otherwise any difference in the velocity fields could be attributed to different regions/densities of the core probed by the two molecules  \citep{Tassis2012}. 3) Radiative transfer effects, the geometry of the B-field, and the morphology of the core can all hinder the detection \citep{Tritsis2023}.

Two observational signatures of ambipolar diffusion  have been proposed: 1) a difference in the systemic velocity between ion and neutral molecules tracing {\rev the same gas \citep[e.g.,][]{Ciolek2000,Tritsis2023}, } 2) broader linewidth of neutrals, which are drifting, compared to the ions attached to the field  
\citep[e.g.,][]{Houde2000}. 
The intensity \citep{Pineda2024} and velocity \citep{Tang2018} power spectra of interstellar clouds may also provide insights on interstellar turbulence dissipation and decoupling of ions and neutrals.

Very few observations have reported detections of ion-neutral linewidth and/or velocity differences. Three papers from the same team, reported broader neutral linewidths
compared to the ion linewidths towards the M17 \citep{Li2008}, DR21(OH) \citep{Hezareh2010}, and NGC 6334 \citep{Tang2018} high-mass star forming regions, using the HCN-HCO$^+(4-3)$ and H$^{13}$CN-H$^{13}$CO$^+(4-3)$ neutral-ion pairs. 
The observed ion-neutral linewidth difference was used to estimate the plane-of-the-sky B-field strength towards the star forming regions.  These three works did not find ion-neutral velocity differences. 
More recently, 
\citet{Pineda2021} analyzed the Barnard 5 star forming region using N$_2$H$^+$ and NH$_3$ and found a  mean ion-neutral velocity difference of $\sim0.05\,$km/s, which they did not discuss as being a signature of ambipolar diffusion. 
The critical densities of N$_2$H$^+(1-0)$ and NH$_3(1,1)$ at 10\,K are $6\times10^4\,$cm$^{-3}$ and $2\times10^3\,$cm$^{-3}$, respectively \citep{Shirley2015}, making these two molecules and transitions likely not tracing the same gas. 
As opposed to the previous observational results, \citet{Pineda2021}  detected  larger linewidth in the N$_2$H$^+$ ion compared to the NH$_3$ neutral molecule, with a difference of about $\sim0.015\,$km/s (for a spectral resolution of $\sim0.05\,$km/s). They proposed this velocity dispersion difference to indicate the penetration of magnetohydrodynamic waves within the densest regions of Barnard 5  inducing turbulent oscillations of the ions, which are  better coupled to the B-field as opposed to the neutrals. 
On smaller scales, ALMA observations of H$^{13}$CO$^+(3-2)$ and C$^{18}$O$(2-1)$ were used  to put an upper limit on the ion-neutral drift velocity of $\sim0.35\,$km/s  at 100 au scales towards  the  Class 0 protostar B335 \citep{yen2018}.

Detecting the ambipolar diffusion process in  prestellar cores, before the onset of protostellar feedback, has important implication in our understanding of the evolution of the B-field flux during the prestellar core collapse.
In addition, magnetic fields are important for transporting angular momentum from the collapsing cores, and play a significant role in the physics of bipolar outflows and jets, and the formation of protoplanetary disks that accompany protostar formation \citep{Machida2008,Pudritz2019}. 
Quantifying observationally the ion-neutral drift velocity and the diffusion of the B-field flux via ambipolar diffusion is therefore crucial to study the collapse and evolution of prestellar cores.

In this paper we present new IRAM-30m telescope observations of deuterated N-bearing  ion and neutral molecules towards the L1544 prestellar core. 
{\rev We have selected  L1544, which is a well studied prototypical prestellar core %
\citep[e.g.,][and references therein]{Spezzano2017,Redaelli2022,Caselli2022,Jensen2023}. %
  We justify the choice of the selected ion-neutral pair in Section\,\ref{mol_selection} and 
describe} the  observations  in Section\,\ref{obs}. The spectral fitting method is described in  Section\,\ref{SpecFit}. 
The results are presented in Section\,\ref{results} and discussed in Section\,\ref{disscussion}. 
Appendices\,\ref{App1}, \ref{Sect:totVel}, \ref{App_comp}, and \ref{model} complement the main text sections with additional details.
Section\,\ref{conc} summarizes and concludes the paper.

\section{Selection of the ion-neutral pair}\label{mol_selection}

Deuterated N-bearing molecules %
are late-type species formed in cold ($\lesssim 15$\,K) and dense  ($\gtrsim10^5$\,cm$^{-3}$) gas,
where their formation is boosted by the catastrophic CO freeze-out, which happens in similar physical conditions  \citep{Caselli1999,Tine2000,ObergBergin2021}. %
N-bearing species, 
{\rev especially NH$_3$ and N$_2$H$^+$}  (and their deuterated isotopologues) 
trace well the central regions of prestellar cores, due to  the fact that nitrogen maintains a significant abundance also at densities where the CO is almost completely depleted onto dust grains \citep[e.g.,][] {Caselli2002,Maret2006,Hily-Blant2010,Johnstone2010}.
{\rev In addition, NH$_3$ and N$_2$H$^+$ are both formed from molecular nitrogen in the dense gas phase 
\citep[][]{Hily-Blant2013,LeGal2014},  %
implying they both form above similar density thresholds in the  inner parts of the cores with  Bonnor-Ebert like density profiles  \citep[][]{Bonnor1956,Aikawa2004,Keto2010}. 
{\revn Observations of NH$_3$ and N$_2$H$^+$  towards a statistical sample of  dense cores confirm the tight correlation between both the velocity centroids and linewidths  of 
 these two N-bearing species, as well as the abundance ratio between NH$_3$ and N$_2$H$^+$ across all studied  cores
 \citep[][{\revl cf. also \citealt{Chitsazzadeh2014}}]{Johnstone2010}. The observed  strong correlations between the physical parameters of   NH$_3$ and N$_2$H$^+$ suggest that their formation and destruction mechanisms (as well as those of their deuterated isotopologues) are similar validating  them as powerful tracers of the same gas towards dense cores.} 
Single dish observations have detected a small (a factor of a few) decrease in the abundance of  NH$_3$ and N$_2$H$^+$  towards the inner parts of prestellar cores, as opposed to the  strong (of few orders of magnitude) freeze-out  detected for C/O-bearing species \citep[e.g.,][]{Caselli1999,DiFrancesco2004,Crapsi2005,Pagani2007,Chitsazzadeh2014,Redaelli2019}. 
Recent  high-sensitivity observations have detected depletion in the deuterated \n2dp  and NH$_2$D species at high densities ($>>10^6$\,cm$^{-3}$) and small  ($<<0.1$\,pc) scales  \citep[][]{Redaelli2019,Caselli2022}. 
Despite low {\revn freeze-out} levels, \n2dp  and NH$_2$D  have been shown to be ideal probes to study the velocity and chemical structures of 
prestellar cores at densities $10^{5-6}$\,cm$^{-3}$ and $\sim 0.1$\,pc scales \citep[e.g.,][]{Ceccarelli2014,Redaelli2019,Redaelli2025,Spezzano2025,Caselli2025}. %
Moreover, these deuterated species (similarly to other molecules probing the dense gas within prestellar cores) show very narrow linewdiths with subsonic velocity dispersions ($<0.2$\,km/s), making them ideal tracers to detect  small velocity shifts  between ion and neutral molecules expected in the presence of ambipolar diffusion.} 

{\rev For this work, we thus have chosen} to compare the velocities of \n2dp$(2-1)$  and para-NH$_2$D$(1_{11}-1_{01})$, which have similar critical densities within a factor of 3, with  $4\times10^5$\,cm$^{-3}$ and $1.4\times10^5$\,cm$^{-3}$ (computed at 10\,K), respectively, making them highly likely to be tracing similar densities and regions  within the $\sim0.1\,$pc scale of  prestellar cores. 
{\rev This is supported by the integrated intensity of these two molecules showing their similar {\revn spatial} distribution within the $\sim0.1\,$pc scale of the L1544 prestellar core (cf. the integrated intensity maps in Appendix\,\ref{App1}).} 

\section{Observations}\label{obs}

The observations were carried out with the IRAM-30m telescope at Pico Veleta, Spain. 
The EMIR multiband mm receiver \citep{Carter2012} was used with the VESPA\footnote{https://web-archives.iram.fr/ARN/dec02/node6.html} autocorrelator as backend. We observed with high-spectral resolutions %
corresponding to velocity resolutions of 0.0379\,km/s for \n2dp and 0.0266\,km/s for  \pnh2d at the rest (unsplit) frequencies   of 
 154217.0112\,MHz %
for \n2dp$(J=2-1)$    and  110153.5940\,MHz
for  para-NH$_2$D$(J_{Ka,Kc} =1_{11}-1_{01})$ %
\citep[cf. e.g.,][]{Dore2004,Daniel2016}. 
The $V_{\rm lsr}$ for both cubes was set to 7.2\,km/s. 
For the rest of the paper we will refer to \n2dp$(2-1)$  and para-NH$_2$D$(1_{11}-1_{01})$, as \n2dp  and \pnh2d, respectively.

The on-the-fly mapping mode was used to observe maps centered towards the dust peak of the L1544 core. The central coordinates of our maps are  RA$_{J2000.0}= 05^h04^m16.539^s$ and  Dec$_{J2000.0}=25^\circ10'47\parcs54$. The position-switching mode was used with a reference position offseted by $\sim10'$ from the target position. The calibration was achieved by measuring the emission from the sky, an ambient load and a cold load every $\sim15\,$min. The telescope pointing was corrected every $\sim2\,$h. We reduced the data with the GILDAS/CLASS software package (the Grenoble Image and Line Data Analysis System, a software  developed by IRAM\footnote{http://www.iram.fr/IRAMFR/GILDAS}). The  position-position-velocity (PPV) cubes of the observed molecules  are  reprojected onto the same spatial pixel grid at the half-power beam width (HPBW) resolution of 23\parcs5, or 0.019\,pc at the 170\,pc distance of L1544 \citep[cf.][]{Galli2019}, and pixel size of 8\arcs. 
The typical rms of the spectra (in Ta$^*$ scale) are 0.02\,K and 0.05\,K, for \n2dp\ and \pnh2d\, respectively.

\begin{figure*}[ht!]
   \centering
  \resizebox{8.9cm}{!}{\includegraphics[angle=0]{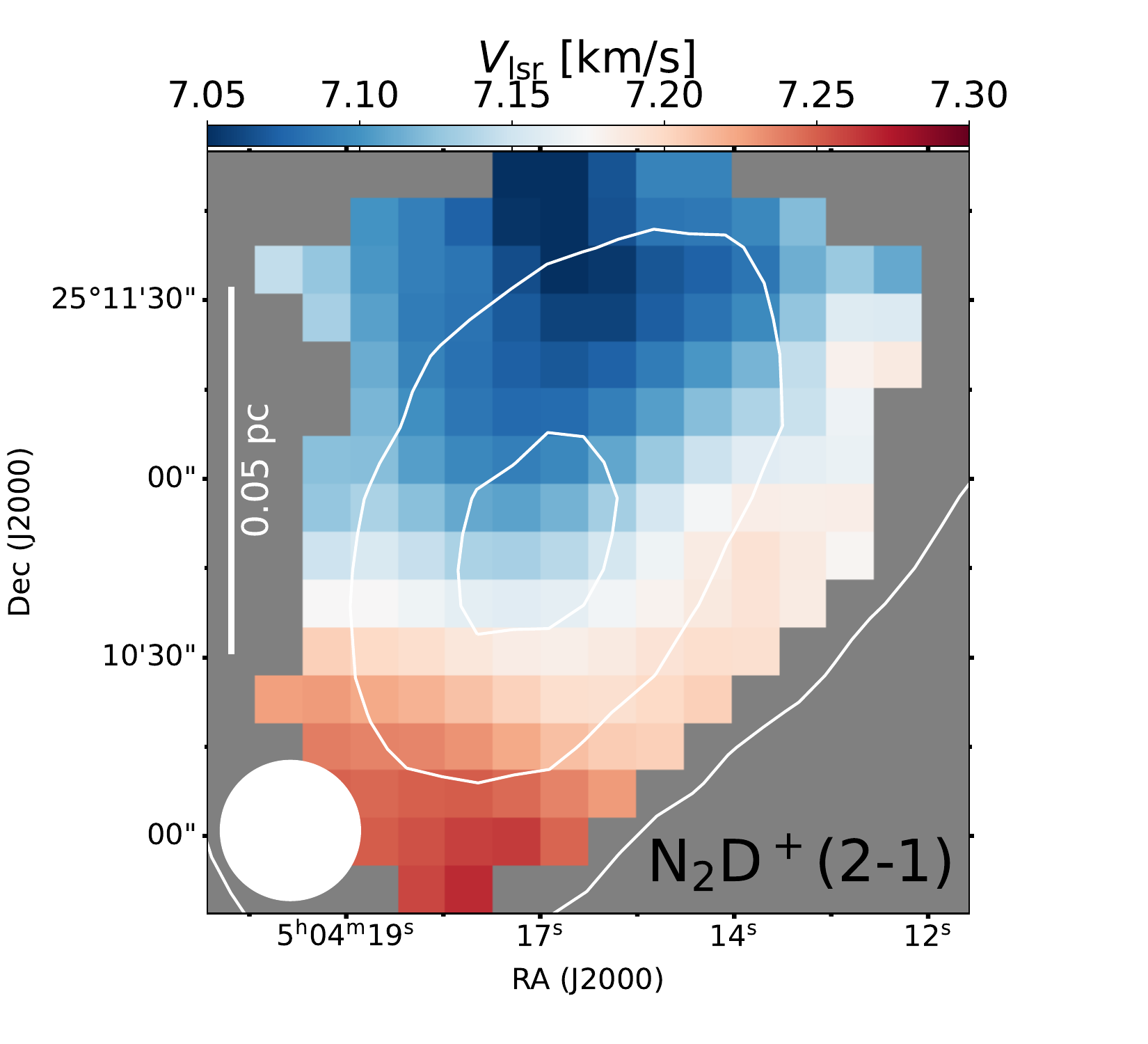}}
\resizebox{8.9cm}{!}{\includegraphics[angle=0]{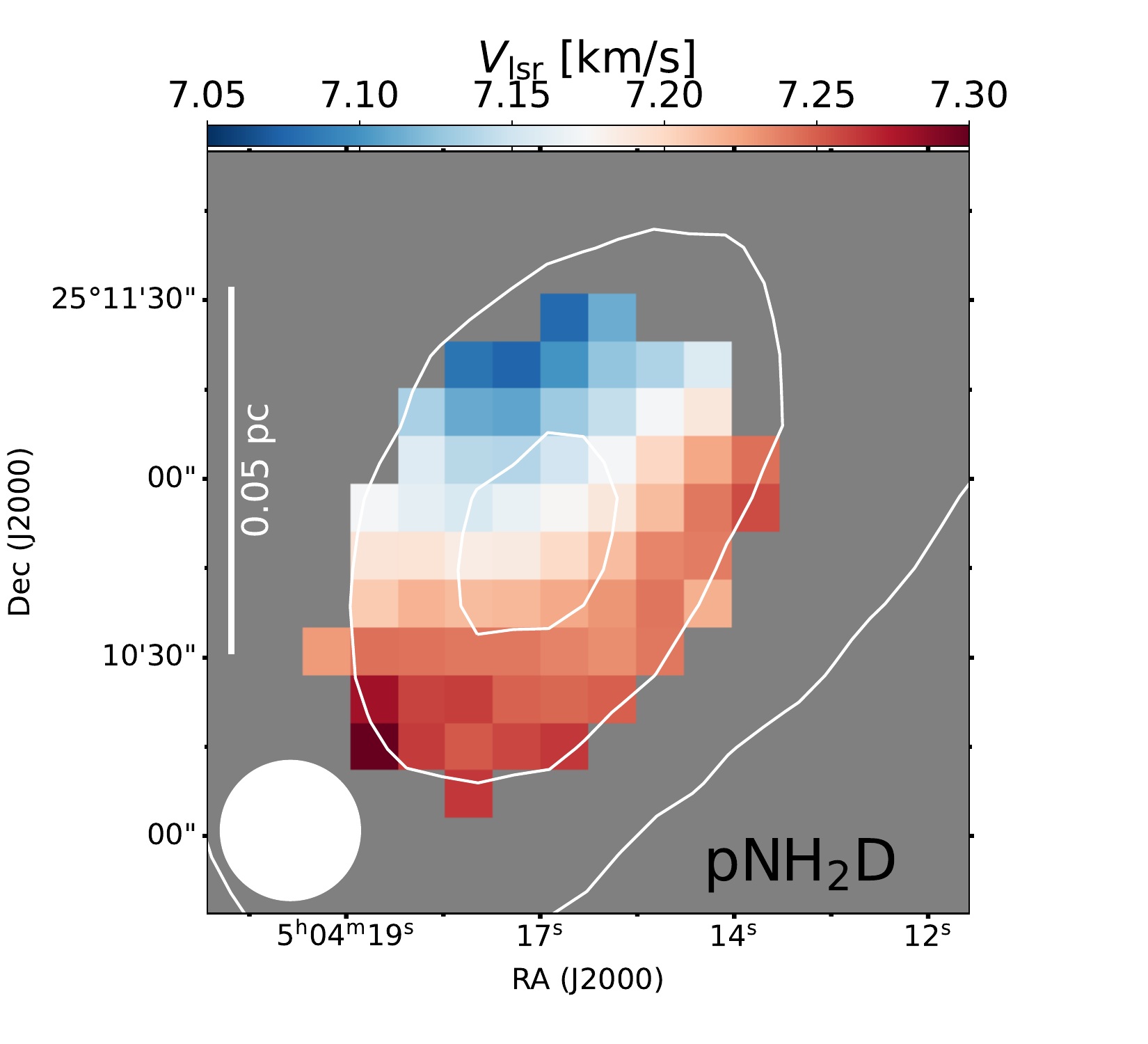}}
  \resizebox{8.9cm}{!}{\includegraphics[angle=0]{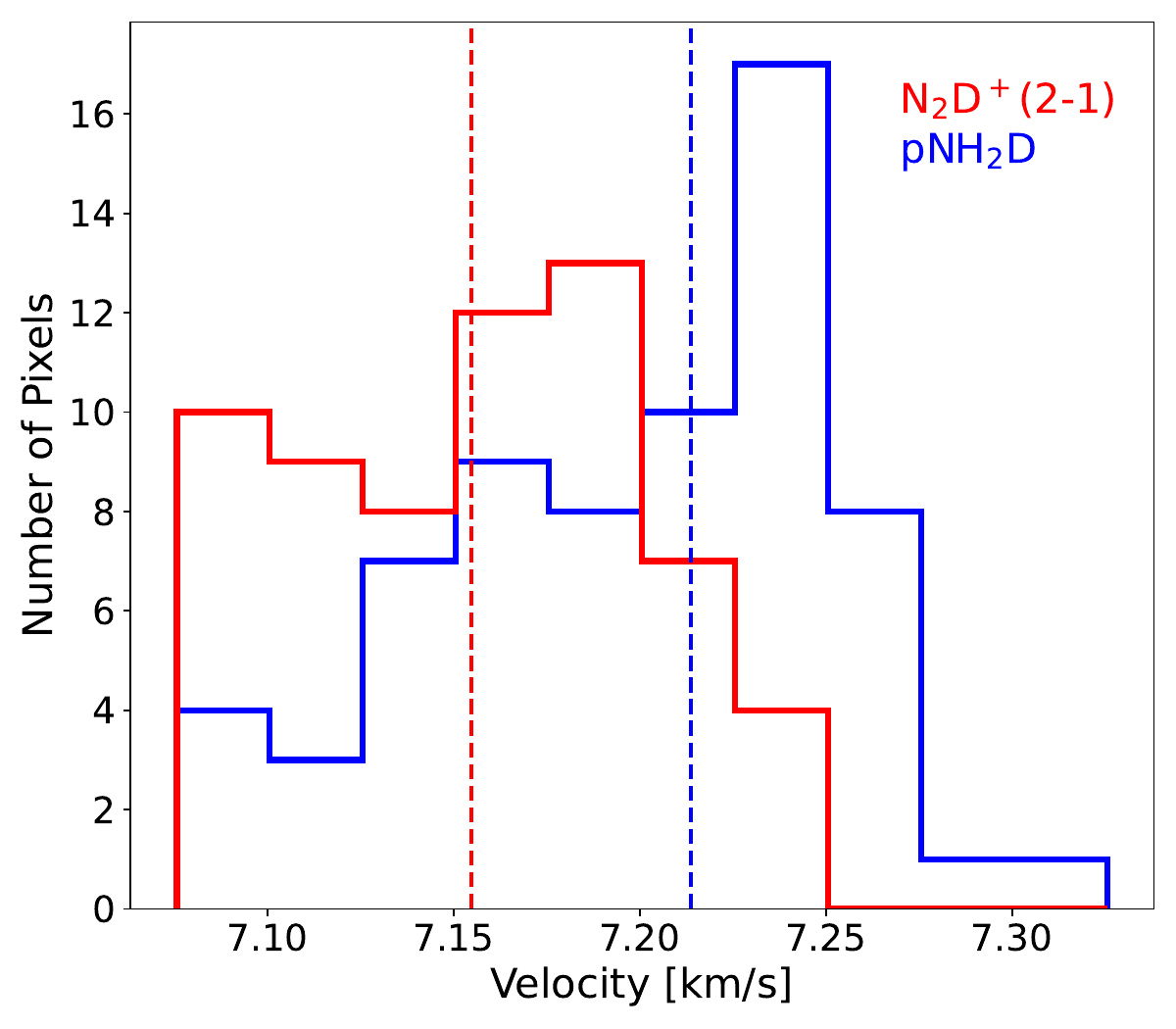}}
\resizebox{8.9cm}{!}{\includegraphics[angle=0]{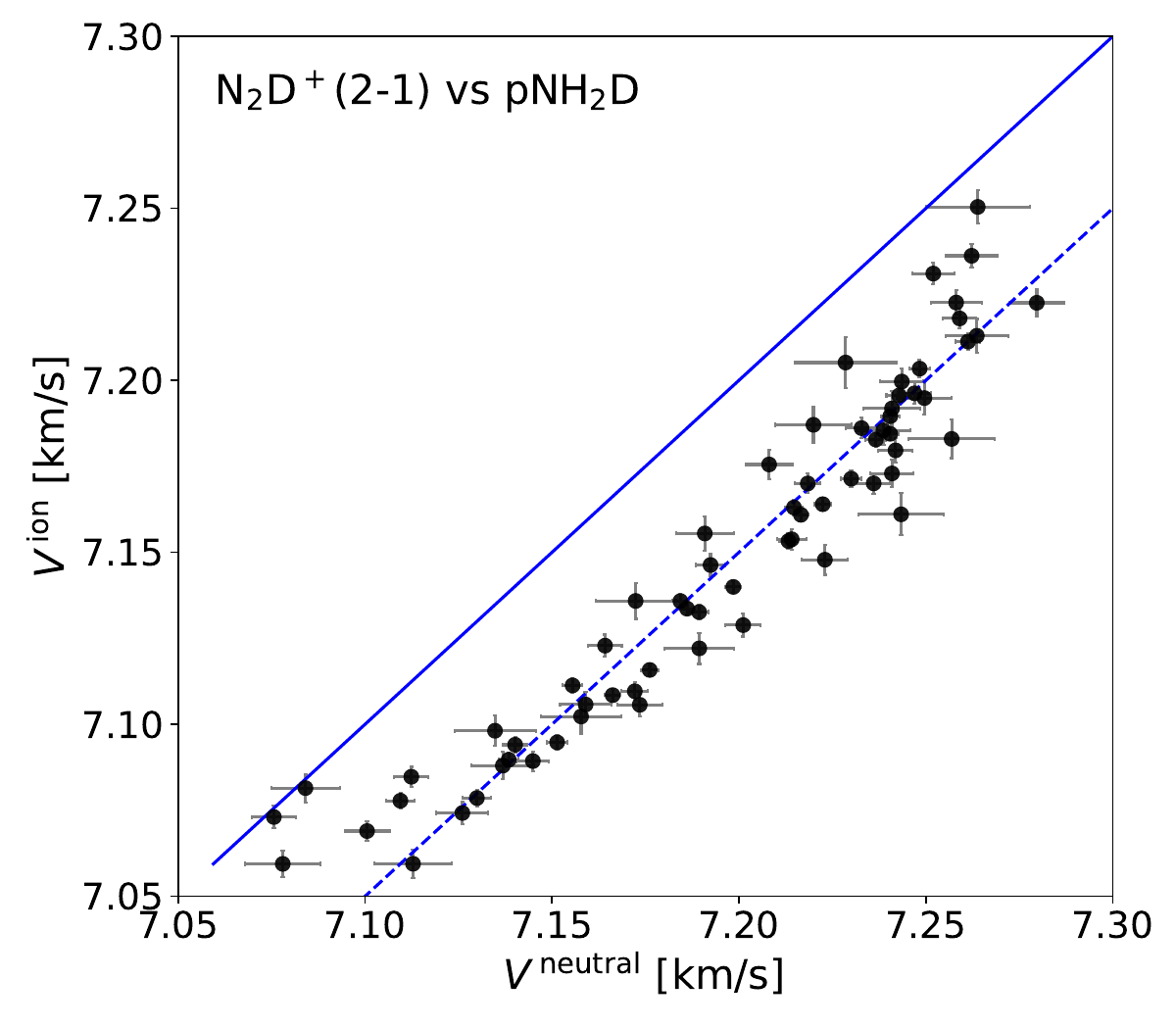}}
  \caption{
   Centroid velocity maps  derived from pixel-per-pixel fitting of the  \n2dp ({\it top-left}) and \pnh2d ({\it top-right}) spectra with $pyspeckit$ (cf. Section\,\ref{SpecFit}). 
      The white contours at $0.8, 1.7,$\,and $2.5\times10^{22}$\,cm$^{-2}$  trace the gas column density  as derived from \herschel\ data \citep[from][]{Spezzano2017} and are the same on both panels. The white filled circles show the 23\parcs5  beam size (0.019\,pc at the 170\,pc distance of L1544) 
     {\rev and the white vertical lines indicate the 0.05\,pc scale.} 
      The excitation temperature maps are shown in Appendix\,\ref{App1}.
The bottom plots show the distribution of the velocity for \n2dp\ in red and \pnh2d\  in blue ({\it bottom-left}) and a scatter plot ({\it bottom-right}). The median values of the distributions indicated by the vertical dashed lines are 7.16\,km/s for  \n2dp\ and 7.21\,km/s for \pnh2d. The oblique solid and dashed blue lines indicate the 1:1 and the  $-0.05$\,km/s relations, respectively. The gray error bars on the scatter plot are the $\pm 1\sigma$ uncertainties on the centroid velocities derived from the $pyspeckit$ fit. The mean values of these uncertainties are 0.003 and 0.006\,km/s for the \n2dp\ and \pnh2d\ data points, respectively (cf. the histograms of the uncertainties are presented in Appendix\,\ref{App1}).
  }          
  \label{Fig:VMaps}
\end{figure*}

\section{Spectral fitting}\label{SpecFit}

We used the $pyspeckit$ analysis toolkit \citep{Ginsburg2011,Ginsburg2022AJ} to fit the hyperfine structure of the  \n2dp  and \pnh2d  spectra, with 26 and 35 hyperfines, respectively. $pyspeckit$\footnote{https://github.com/pyspeckit/pyspeckit/tree/master/pyspeckit/spectrum/}
has the most recent libraries of hyperfine frequencies for these two molecules. 
The fitting results from $pyspeckit$ include: 
The excitation temperature  $T_{\rm ex}$, the optical depth $\tau$, the velocity in the local standard of rest (lsr) $V_{\rm lsr}$, the velocity dispersion $\sigma_{\rm v,obs}$, and the associated uncertainties for all these four parameters derived from the chi-square minimization approach \citep{Ginsburg2022}. 
We have validated the fitting results using the hyperfine module (HFS mode) of Gildas/CLASS$^2$. 

{\rev 
The accuracy of the spectroscopy of the  \n2dp\ \citep{Dore2004,Pagani2009} and \pnh2d\ \citep{Daniel2016,Melosso2021} lines reported in  the catalogs, on the order of $0.005-0.008$\,km/s, is smaller by a factor $3-7$ compared to the velocity resolution of our data.
 Consequently, the observed differences between the centroid velocities of the lines (cf. Section\,\ref{results})
are unlikely to 
originate from the considered rest frequencies derived from spectroscopic measurements.} 
We also note that the velocity accuracy is homogeneous for all hyperfine transitions \citep[cf.][]{Ginsburg2022}.

The spectra with peak signal-to-noise ratio smaller than 5 have been excluded from the analysis. 
In Appendix\,\ref{App1}, %
we show  representative  \n2dp\ and \pnh2d\  spectra and the best fit model  towards the peak intensity of the core, i.e., the central pixel of the maps. We also present the derived fitting uncertainties on the velocities.

To compare the velocity dispersions of the two molecules of interest \n2dp\ and \pnh2d, we estimate the total velocity dispersion of the mean free particle. We also derive the  non-thermal velocity dispersion of the gas by subtracting the thermal velocity dispersion from the observed linewidth adopting a gas temperature of 10\,K, which is the mean gas temperature of L1544 at the {\rev observed  $\sim 0.05-0.1\,$pc  scales and  $\sim 10^{5-6}\,$cm$^{-3}$  densities} \citep{Crapsi2007,Spezzano2017}.
We derive maps of   $ \sigma_{\rm v,tot}$ and $\sigma_{\rm v,NT}$ for both \n2dp\ and \pnh2d. 
 The details and results of the calculations are given in Appendix\,\ref{Sect:totVel}.

\begin{figure*}[ht!]
   \centering
  \resizebox{8.9cm}{!}{\includegraphics[angle=0]{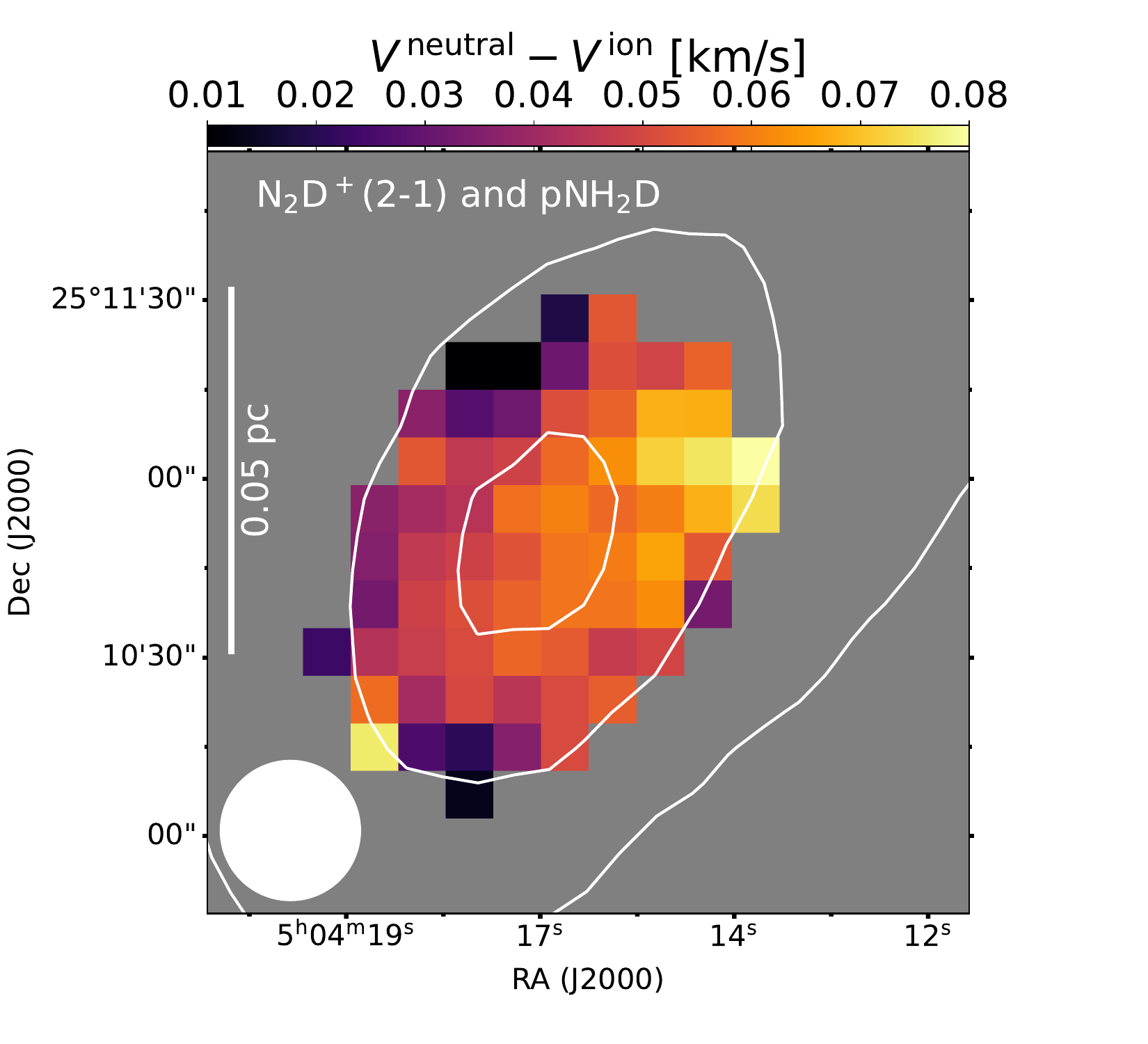}}
\resizebox{8.9cm}{!}{\includegraphics[angle=0]{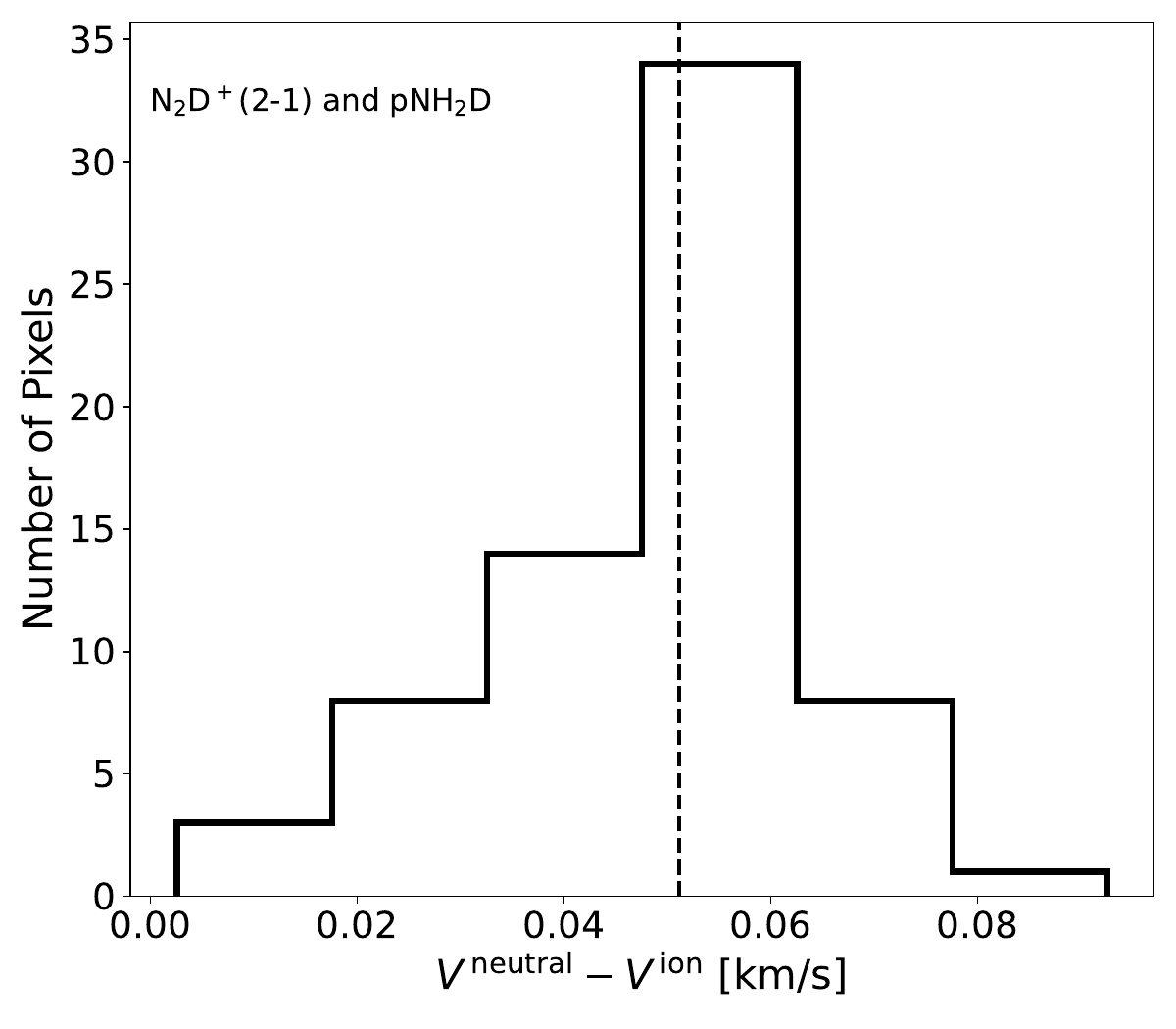}}
  \caption{
 {\it Left:} Map of the  difference between the  neutral \pnh2d and the ion  \n2dp   centroid velocity ($\delta V^{\rm shift}=V^{\rm neutral}-V^{\rm ion}$). The filled white circle shows the 23\parcs5  beam size (0.019\,pc at the 170\,pc distance of L1544) 
 {\rev and the white vertical line indicates the 0.05\,pc scale.} 
 {\it Right:} Histogram of $\delta V^{\rm shift}$. The mean and standard deviation of the  $\delta V^{\rm shift}$ distribution are 0.05\,km/s and 0.02\,km/s, respectively, where the mean uncertainty on $\delta V^{\rm shift}$ is 0.006\,km/s %
 (cf. Appendix\,\ref{App1}).
  }          
  \label{Fig:Vdrift}
\end{figure*}

 \begin{figure}[ht!]
   \centering
  \resizebox{9.cm}{!}{\includegraphics[angle=0]{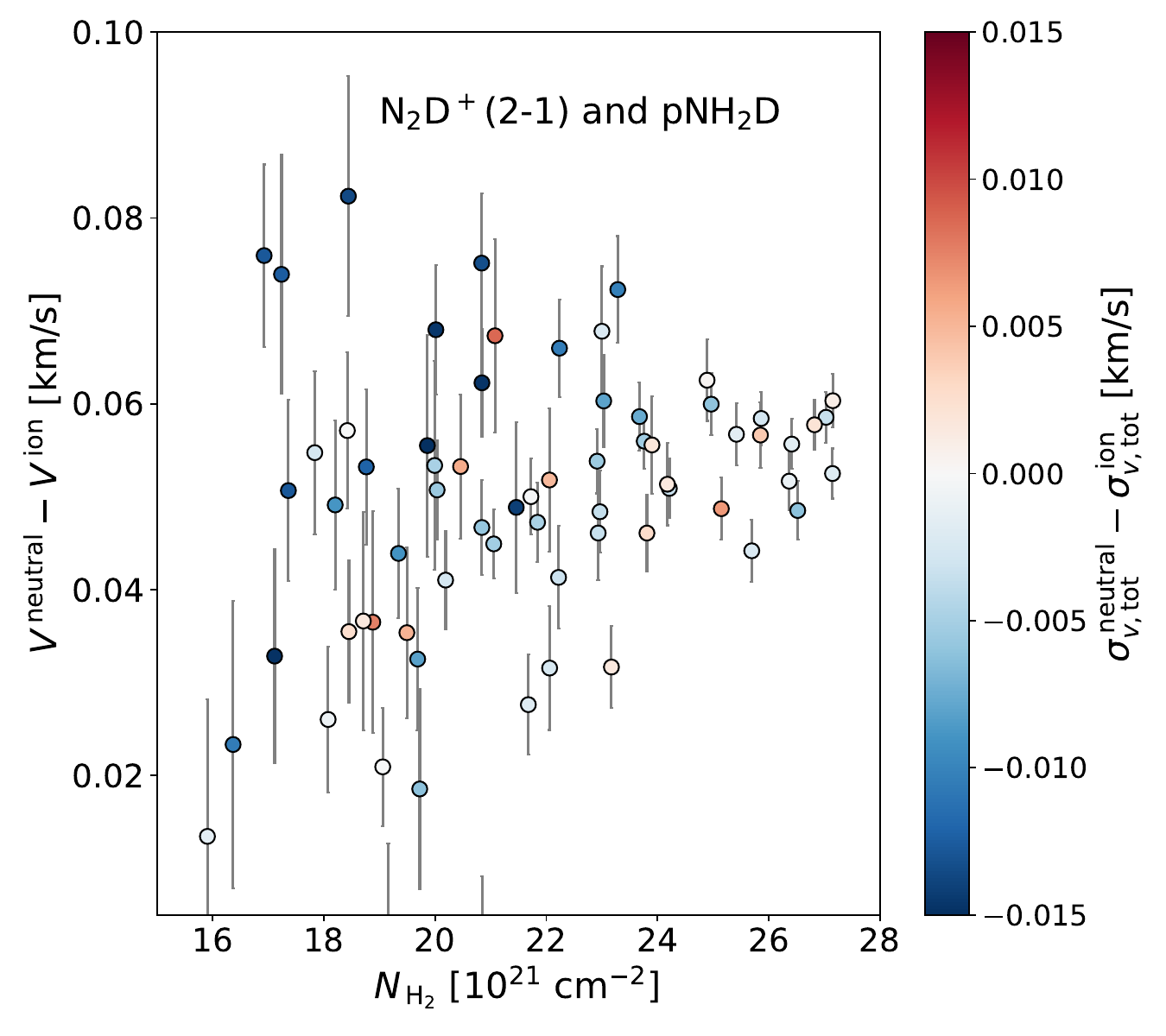}}
  \caption{Scatter plot of  the  difference between the  neutral \pnh2d and the ion  \n2dp  centroid velocity ($\delta V^{\rm shift}=V^{\rm neutral}-V^{\rm ion}$) versus the $N_{\rm H_2}$ column density derived from \herschel\ and reprojected to the same spatial grid as that of the molecular line data. The scatter in $\delta V^{\rm shift}$ is larger for smaller $N_{\rm H_2}$  values. 
 {\rev The color of the data points and the color bar indicate the difference of the ion-neutral total velocity dispersion ($\delta \sigma_{\rm v,tot}$).}
  }          
  \label{Fig:coldensVSveldiff}
\end{figure}

\section{Results: Ion-neutral velocity difference}\label{results}

  The integrated intensity and the excitation temperature  $T_{\rm ex}$ maps of \n2dp\ and \pnh2d\ show the  highly centrally concentrated emission within the $\sim0.05$\,pc diameter of the core (cf. Appendix\,\ref{App1}). 
   {\revn The integrated intensity radial profiles of \n2dp\ and \pnh2d\  are very similar, with only a $\sim20\%$ larger value for \pnh2d\ within the central part of the core. }
  On the $T_{\rm ex}$ maps, 
 \n2dp\ shows a slightly ($\sim25\%$) more extended emission compared to that of \pnh2d\footnote{\revl{The spatial distribution of N$_2$H$^+$ towards prestellar cores also shows a slightly more extended emission compared to that of NH$_3$ \citep[see, e.g.,][]{Tafalla2002}, indicating the possible inheritance of  the spatial distributions of their deuterated species.}}, which is highly concentrated for gas column densities $>1.7\times10^{22}$\,cm$^{-2}$ as derived from \herschel/SPIRE data \citep[][]{Spezzano2016}. {\rev This latter column density corresponds to a volume density of $\sim 10^{5}\,$cm$^{-3}$ %
 assuming a spherical density structure (with a diameter of 0.05\,pc).} 
 The mean and standard deviation of the $T_{\rm ex}^{\rm neutral}/T_{\rm ex}^{\rm ion}$ ratio and the $\tau^{\rm neutral}/\tau^{\rm ion}$ are $1.03\pm0.44$ and  $0.99\pm0.18$ (cf. histograms in Appendix\,\ref{App1})  
 suggesting that these two molecules are  tracing similar optical depth and  densities, making them strong candidates for 
 chemically co-evolving molecules within the core, further supporting  our choice of ion-neutral pair to investigate signatures of ambipolar diffusion towards L1544.
 
Figure\,\ref{Fig:VMaps} presents the velocity structure of L1544 as mapped with  \n2dp and \pnh2d. The spectra with peak signal-to-noise ratio smaller than 5 excluded from the analysis are shown with uniform gray colors.

 {\rev The velocity maps for both \n2dp and \pnh2d, show the presence of  velocity gradients along and across L1544. Such velocity gradients are usually attributed to rotation and infall  \citep[e.g.,][]{Tafalla1998,Ohashi1999,Caselli2002.I,Williams2006}.}
The mean and standard deviation of the  centroid velocity distributions are 7.199\,km/s and 0.054\,km/s for \n2dp\ and 7.151\,km/s and 0.050\,km/s for \pnh2d. The velocity scatter plot on Figure\,\ref{Fig:VMaps} highlights the  systematic velocity shift %
between  \n2dp and \pnh2d. The spatial distribution of the velocity difference is given in the left panel of  Figure\,\ref{Fig:Vdrift}
where larger values are found in the northwest part of the core 
decreasing values from the southeast to the northwest. 
The right panel of Figure\,\ref{Fig:Vdrift}  shows the velocity difference $\delta V^{\rm shift}$  histogram with 
 a mean value of $0.05$\,km/s and a standard deviation of $0.02$\,km/s. %
 The detected ion-neutral mean $\delta V^{\rm shift}$ value is spectrally resolved by a factor of 1.7 and the standard deviation of the distribution is smaller than the ($\sim 0.03\,$km/s)  velocity resolution of our current data.   
 
 {\rev The observed \n2dp and \pnh2d spectral lines are narrow and characterized by  velocity dispersions dominated by thermal motions, with $ \sigma_{\rm v,tot}\sim0.2\,$km/s and $\sigma_{\rm v,NT}\sim0.1\,$km/s (cf. Appendix\,\ref{Sect:totVel}).}
 On average, a difference of $-0.004$\,km/s  is  detected 
 in the total velocity dispersion  of both species with a standard deviation of %
 $0.006$\,km/s (cf. Appendix\,\ref{Sect:totVel}). 
 This measured difference of the ion-neutral velocity dispersion  is an order of magnitude smaller than the velocity resolution of our data ($\sim 0.03\,$km/s).

Figure\,\ref{Fig:coldensVSveldiff} presents a scatter plot of the ion-neutral velocity difference as a function of  the column density derived from $Herschel$ data. For column densities $>2.4\times10^{22}$\,cm$^{-2}$, towards the center %
of the core,  $\delta V^{\rm shift}\sim0.05$\,km/s. For these data points  $\delta \sigma_{\rm v,tot}\sim0.005$\,km/s indicating $ \sigma_{\rm v,tot}^{\rm neutral}>\sigma_{v,tot}^{\rm ion}$. %
For column densities $<2.4\times10^{22}$\,cm$^{-2}$ the scatter of both $\delta V$ and $\delta \sigma_{\rm v,tot}$ is larger. 

The spatial resolution of our data ($\sim 0.019\,$pc) and {\rev the narrow column density range (only a factor of 1.8) probed by our observations are} a limitation to further discuss the  spatial variations in the ion-neutral velocity difference {\rev and velocity dispersion}. 
Higher spatial resolution observations {\rev spanning a broader  column density dynamic range} are needed to  resolve  the core and study the spatial velocity variations.

\section{Discussion}\label{disscussion} 

\subsection{Interpretation of the observed velocity difference}\label{diss1}

To fully characterize the initial condition of star formation and the onset of collapse of magnetized prestellar cores  it is essential to describe the physical processes and the timescales governing the diffusion of the magnetic field. 
The magnetic flux is 
{\rev reduced within the core} 
due to the partial decoupling of the charged and neutral particles allowing the neutrals to slip through the B-field lines while the charged particles feeling the Lorentz force {\rev remain} attached to the B-field. This decoupled infall motions result in a drift velocity $v_{\rm drift}=|V^{\rm neutral}-V^{\rm ion}|$, where 
\begin{equation}
    v_{\rm drift}\sim\frac{\eta_{\rm AD}}{l}, 
\end{equation}
with $\eta_{\rm AD}$ the ambipolar resistivity and $l$ the characteristic scale of B-field fluctuations. %
 On microscopic level the 
  ambipolar resistivity is a function of the total density ($\rho$), the magnetic field strength ($B$), and the ionization degree ($x(e)$)  
  of the core.  When the adsorption of charged particles onto dust grains is negligible,  the 
  ambipolar resistivity is expressed as
  \begin{equation}
   \eta_{\rm AD}(\rho,B,x(e))\propto \frac{B^2}{\rho^{2}x(e)}\label{eq:etaAD} 
   \end{equation}
  The ionization degree within dense cores $x(e)$, without internal protostellar source, is controlled by the interstellar cosmic-ray ionization rate, the gas-phase recombination, and the dust grain size distribution \citep{McKee1989,Caselli1998,Ivlev2015}.  The cosmic-ray ionization rate in prestellar cores, including L1544, is observationally estimated to be $\zeta_{\rm H_2}\sim 10^{-17}\,$s$^{-1}$ \citep{Redaelli2021,Redaelli2025}. Equation\,\ref{eq:etaAD} considers only gas-phase recombination. When dust grains are present, the adsorption of charged particles onto the dust grains could also play a role in which case, the expression of $\eta_{\rm AD}$ differs from that of    
  Equation\,\ref{eq:etaAD}.

  {\rev
Several dynamical models ranging from early studies \citep[e.g.,][]{Basu1995,Tassis2007} to more recent 3D simulations \citep[e.g.,][]{Tritsis2022,Tritsis2025} have explored the impact of ambipolar diffusion on core collapse.}  
  {\rev While these latter studies 
investigated ion-neutral drift velocities 
without 
invoking the evolution of the dust size distribution, 
the strong dependence of $\eta_{\rm AD}$ (and hence $v_{\rm drift}$)  on the abundance of small grains in the core centers has been   discussed in a number of recent works \citep[e.g.,][]{Zhao2021,Tsukamoto2022}. }
 \citet[][]{Silsbee2020} have  investigated the differential grain drift due to ambipolar diffusion affecting the dust coupling to the magnetic field and to the neutral gas.  They showed that small grains (mostly negatively charged), which stay coupled to the magnetic field are efficiently depleted within the core center due to a rapid coagulation with larger grains, drifting
 through the magnetic field towards the core center due to infall motions. They calculated an  ion-neutral velocity drift of $\sim0.05\,$km/s for L1544. %

  In a dedicated recent study, \citet[][]{Fukihara2026} calculated the time evolution of $\eta_{\rm AD}$ at dense core densities of $10^6$\,cm$^{-3}$ as grains growth via {\rev accretion and coagulation\footnote{\citet{Fukihara2026} {\rev did not consider fragmentation due to grain-grain collisions because the relative grain velocity in their calculation is below  ($<100$\,m/s) the fragmentation limit  \citep[as shown in][]{Kawasaki2022}. See also \citet{Lebreuilly2023} for a discussion on the impact of fragmentation on dust growth.}}}. They demonstrate that within several free-fall times, %
  dust grains grow  sufficiently contributing to the increase of $ \eta_{\rm AD}$ resulting in  a more efficient decoupling between charged and neutral particles.
  They demonstrate
  that $v_{\rm drift}\sim0.05-0.10\,$km/s due to ambipolar diffusion are expected at core densities of $10^6$\,cm$^{-3}$ with magnetic field fluctuations on the order of the  {\rev Jeans length  %
  (corresponding to the characteristic scale of density fluctuations in self-gravitating  cores {\revn in frozen-in conditions})}
  when the following three conditions are satisfied: 
  1) magnetic field strengths are $100-300\,\mu$G, 2) $\zeta_{\rm H_2}\sim 10^{-17}\,$s$^{-1}$, and 3) the very small $<0.02\,\mu$m grains (VSG) are depleted due to dust growth. 
  They show that when  the above conditions are not satisfied, the resulting  $v_{\rm drift}$ are smaller ($\sim0.001\,$km/s).
  VSG are found to be efficiently depleted through accretion within one free-fall time $\sim10^4$\,yr for densities of $10^6$\,cm$^{-3}$ followed by the growth of grains via coagulation reaching maximum sizes of $\sim10\,\mu$m within $\sim10$ free-fall times  \citep[][see also \citealp{Ormel2009}]{Fukihara2026}.  

While dust grain growth from observations has not yet been well constrained and still debated, multiple observational results based on different methods strongly indicate the growth of dust grains  from  the diffuse ISM to the dense prestellar cores  \citep[see, e.g., a recent review by][]{Maury2022}: 1) The observed mid-IR scattered light towards starless cores \citep[coreshine, e.g.,][]{Pagani2010}, 2) the change of the dust emissivity index derived from submillimeter dust continuum observations \citep[e.g.,][]{Bracco2017,Galametz2019,Chacon-Tanarro2019}, 3) the observed $2-5\%$ polarization fractions in the submillimeter \citep[e.g.,][]{Valdivia2019,LeGouellec2020}, all require micron-sized grains to account for the observation. 4) Recent spectroscopic  observations of interstellar icy grains with the JWST reveal that grains reach micrometer sizes before the protostellar phase \citep[e.g.,][]{Dartois2024}.

{\rev As presented in Section\,\ref{results}, we detect a $0.05\,$km/s  velocity shift between \n2dp and \pnh2d towards the inner $\sim0.05$\,pc scale of L1544 (Figure\,\ref{Fig:VMaps}). 
We propose that}
the theoretically predicted ion-neutral drift velocities at dense core densities including dust grain growth on the order of $0.05-0.1\,$km/s suggest that 
the observed $0.05\,$km/s velocity difference between \n2dp and \pnh2d towards L1544 may be tracing the decoupling between ions and neutrals due to ambipolar diffusion. 
 We detect the  $0.05\,$km/s velocity shift  towards the inner $\sim0.05\,$pc of the core. {\revn This $\sim0.05\,$pc scale}  is on the order of the  Jeans length $\sim0.06$\,pc (at the density of $10^5$\,cm$^{-3}$ and temperature of 10\,K, cf. Section\,\ref{results}), suggesting  
that the characteristic scale of B-field fluctuations/bending due to gravity within the  core may be on the order of the Jeans length {\revn as suggested by \citet[][]{Fukihara2026} in the specific conditions of spherically collapsing self-gravitating gas.
Observational studies of the ion and neutral velocity power spectra have found 
ambipolar diffusion length scales 
within the range of  $\sim2$\,mpc \citep{Li2008} and $\sim20-30$\,mpc \citep{Pineda2024,Houde2009,Houde2016}. These latter length scales estimated from the analysis of cloud scale data are smaller than the above mentioned Jeans length for L1544 at the density of $10^5$\,cm$^{-3}$. 
High-angular resolution observations of the velocity and magnetic field structures are required to describe the characteristic scale of magnetic field fluctuations and its origin within L1544. 
}

An important implication in confirming the origin of the observed velocity difference resulting from ambipolar diffusion helps us constrain the total 3D magnetic field strength towards the $\sim0.05\,$pc  of L1544 to be on the order of $100-300\,\mu$G. This predicted range of the magnetic field strength  is compatible with the $140\,\mu$G strength estimated 
from dust polarized emission observations at $850\,\mu$m with the JCMT/SCUPOL  \citep[][see also \citealp{Clemens2016}]{Crutcher2004}. 

\subsection{Projection effects}\label{diss3}

The theoretical studies \citep[][]{Silsbee2020,Fukihara2026} discussed in Section\,\ref{diss1} obtained 3-dimensional (3D) velocity drifts $v_{\rm drift}$ on the order of $0.05$\,km/s for reasonable assumptions for dense core physical parameters (density, magnetic field, ionization), assuming a simplified magnetic field geometry, and also do not consider  asymmetries in the core properties.

Because the observations trace the line-of-sight (LOS) velocity of the gas, both the physical geometry of the core and the spatial structure of the infall/contraction\footnote{
The dedicated model of \citet{Keto2010} reproducing the observed molecular line profiles  of the L1544 prestellar core   demonstrated that L1544 is quasi-statically contracting and 
is not undergoing a free-fall infall (probably due to the magnetic pressure opposing gravitational free-fall). In this section, we use "infall" to describe the kinematics of cores, where the infall motions correspond to quasi-static contraction for L1544.}
motions strongly influence the observed velocity pattern
of $V_{\rm lsr}$. {\revn In the absence of turbulence,} 1) when the core has a uniform spherical density structure  undergoing a symmetric  infall, with or without ion-neutral drift velocities, no LOS velocity gradients nor drift velocities can be detected. {\revn In the presence of (anisotropic) turbulence, however, the symmetry of the velocity field is broken and both  velocity gradients and velocity drifts can be detected.} 2) When the core has a flatten ellipsoidal density structure, velocity gradients tracing infall are observable and the presence of the LOS ion-neutral velocity drift is characterized by a sign inversion {\rev between the redshifted and blueshifted velocity gradients of the infalling core}. 3) When the core is a titled and bent ellipsoid,  velocity gradients are observed and the LOS ion-neutral velocity drift has the same sign. Our observations of L1544 show a velocity gradient with a drift velocity that does not change sign, suggesting that L1544 may be well described by a bent ellipsoid with asymmetric  velocities tracing infall/contraction and accretion. The description of a bent-ellipsoid is compatible with the observed maps of L1544 and  with multiple  studies that found asymmetric accretion flows suggested by the observed velocity fields and shock signatures  \citep{ShinyoungKim2022,Lin2022,Giers2025}. 

For increasing densities towards  the core center, $v_{\rm drift}$ is expected to {\revn decrease   \citep[see Equation\,\ref{eq:etaAD} and cf. also][]{Mouschovias1991}.} 
 Our observations probe a column density range of a factor of 1.8 (between $\sim 1.6$ to 2.8$\,\times10^{22}$\,cm$^{-2}$). The velocity drift on such a small range is expected to change on the order {\revn of $40\%$  
 {\revl (assuming $B\propto \rho^{1/2}$)\footnote{
 {\revl This relation is compatible with  observational results estimating the spatial distribution of the B-field strength towards individual filaments and cores at a given evolutionary stage  \citep[][]{Kandori2018,Hwang2021,LeNgan2024}.}}},
 which is on the order of the observed  $v_{\rm drift}$ scatter (Figure\,\ref{Fig:coldensVSveldiff}). The observed scatter, which  may result from a more complex velocity field and affected by projection effects, may hide the expected decrease of $v_{\rm drift}$ with increasing column density. 
{\rev In particular, in their non-ideal MHD simulations, \citet{Tritsis2025} found that at an evolved stage of a turbulent core collapse, the twisting of the B-field lines near the core midplane  induces a fraction of the $v_{\rm drift}$  vectors to point outward. Such a complex velocity  structure would manifest as positive and negative $v_{\rm drift}$  gradients on either side of the core midplane.    
To better constrain the origin of the observed  $v_{\rm drift}$ and its scatter, higher angular (and spectral) resolution data are required to resolve the velocity distribution  and probe a broader column density range within the core. }

{\rev As vectors of the ion-neutral drift velocity are perpendicular to the local B-field, observations of the B-field structure derived from dust polarized emission towards  L1544 would provide invaluable hint on the pinching of the B-field lines and thus on the ongoing ambipolar diffusion process. 
 \citet{Ciolek2000} suggested a model for the mean B-field in L1544 close to the plane-of-the-sky (POS), while recent polarization observations (Kim et al., submitted)  favor a more inclined configuration of the B-field. 
 {\revn In the absence of turbulence,} for an axisymmetric B-field configuration, a more  inclined B-field with respect to the POS will decrease the measured $v_{\rm drift}$ along the LOS. On the other hand,  {\revl  
 an inclined B-field configuration} coupled with the presence of asymmetries in both the density and velocity structures will break the symmetry of the integrated emission allowing the detection of $v_{\rm drift}$ (cf. Section\,\ref{diss4}).
 The presence of asymmetries in the density, B-field, and velocity structures combined with dust growth (cf. Section\,\ref{diss1}), are jointly needed to interpret  the observed velocity difference between \n2dp and \pnh2d towards L1544 as a  drift due to ambipolar diffusion. 
 
 The broadening of the linewidth of neutrals compared to that of ions has been suggested as a signature of ongoing ambipolar diffusion \citep[e.g.,][see also Section\,\ref{intro}]{Li2008}. However, these latter studies have been targeting (optically thick) molecular lines (e.g., HCO$^+$, CO) with supersonic linewidths tracing lower density gas  in the turbulent environments surrounding  star-forming cores (in high-mass star-forming regions). Disentangling between various physical mechanisms (interstellar turbulence, impact of stellar feedback, ambipolar diffusion, radiative transfer effects, opacity effects) that may contribute to the observed differences between the linewidths of these different molecules may be challenging. 
{\revn In contrast to previous studies in supersonic environments,}
 in our observations towards the central part of L1544,
 the  (non-thermal) velocity dispersions of \n2dp and \pnh2d are subsonic (cf. Appendix\,\ref{Sect:totVel}) and the  contraction motions within the $\sim0.05\,$pc L1544 core are also subsonic. 
 Consequently, the subsonic inward motion of the neutrals  decoupled from the B-field and the ions may not have a strong impact in broadening the linewidth of \pnh2d, hence the observed similar linewidth for \n2dp and \pnh2d.}
 {\revn In addition,} the LOS integration and projection effects may also impact the observed linewidth \citep{Priestley2022}, {\revn making the linewidth difference between ions and neutrals not a reliable diagnostic when the magnetized object has some inclination. In particular, for L1544, the detection of the velocity difference between \n2dp and \pnh2d combined with the absence of significant difference of their linewidth would be compatible with an inclined configuration of the system. 
 } 

  To better {\revn constrain the geometric limitations of the observations (velocity centroid and linewidth maps) and} understand the origin of  the observed  velocity and linewidth difference between \n2dp and \pnh2d towards L1544  a quantitative comparison  with  synthetic maps {\rev for various projection angles} derived from  non-ideal MHD simulations including ambipolar diffusion and grain growth is required. Such a study will be presented in a future publication {\rev (Misugi et al., in prep.)}. 

\subsection{Assessing the co-spatiality of the \n2dp and \pnh2d emission }\label{diss2}

We have chosen the N-bearing deuterated molecules \n2dp and \pnh2d to investigate the ion-neutral velocity difference because of their similar critical densities within a factor of 3. The observed ratio of  the fitted $T_{\rm ex}$ and $\tau$ values close to 1 (see Appendix\,\ref{App1}) suggesting that these molecules may be tracing the same densities supporting our choice of selecting them for the analysis aiming at studying the ion-neutral velocity drift due to ambipolar diffusion. The spatial distributions are also very similar (see Fig.\,\ref{Fig:TexMaps}), although higher-angular resolution observations are needed to better resolve the core structure.
Interestingly, \n2dp, which has a critical density of $4\times10^5\,$cm$^{-3}$, 2.8 times larger than the critical density of \pnh2d shows a slightly more extended emission (see Fig.\,\ref{Fig:TexMaps}), which may contribute to broadening the linewidth of the \n2dp  masking the possible detection of  larger neutral linewidth compared to the ion linewidth as proposed by \citet{Li2008}. 
 Radiative transfer effects, line-of-sight integration, {\rev projection effects, and density/velocity asymmetries} may also hinder any possible detection of linewidth differences. 

To better understand the origin of the observed velocity difference, we also compare, in Appendix\,\ref{App_comp}, %
the $V_{\rm lsr}$ of  \n2dp$(2-1)$  and \pnh2d with that of \n2dp$(1-0)$  obtained with the IRAM-30m telescope. This latter ion has a critical density of $8\times10^4\,$cm$^{-3}$ at 10\,K, a factor of 1.7 and 5 lower  than \pnh2d and \n2dp(2-1), respectively. For this comparison, we smooth the three data cubes to  the coarser 34" angular  resolution of  \n2dp(1-0).  The spectral resolution of \n2dp(1-0) is 0.076\,km/s.  We do not resample the three data cubes spectrally, but correct for the channel response  (see Appendix\,\ref{App_comp}). %
We confirm that spectrally smoothing the data to 2 and 3 times coarser resolution does not change the fitted value of the  $V_{\rm lsr}$ and broadens the linewidth minimally ($\sim4\%$). 
We find a mean velocity difference between \pnh2d and \n2dp$(1-0)$  of 0.013\,km/s, which is smaller than the spectral resolution of the data (cf. Appendix\,\ref{App_comp}). 
The distributions of the $T_{\rm ex}$ and $\tau$ ratios between the ion and neutral are broader when the ion is \n2dp$(1-0)$ compared to those when the ion is \n2dp$(2-1)$, suggesting that the \n2dp$(2-1)$ emission is more closely correlated to that of   \pnh2d.

\subsection{Analytical model of an infalling/contracting core with  drift velocity}\label{diss4}

The observed velocity difference may, however, be a combination of ion-neutral velocity drifts and different infall motions of slightly different gas density regimes probed by the chosen  ion-neutral pair. 
 To test whether the observations could be reproduced with a model of an infalling-rotating core with a drift velocity between ions and neutrals, we designed an analytical model of an ellipsoidal core with parametrized geometry, and parametrized density and velocity profiles. The 3D core models are then integrated along the lines-of-sight to derive synthetic spectral maps (i.e., PPV cubes). Thanks to a JAX-based architecture \citep{jax2018github}, the model is completely differentiable with respect to the input parameters. With this approach, the derived synthetic PPV cubes 
are used to optimize the parameters and minimize the difference with the observed PPV cubes.
Further details of the model are described in Appendix\,\ref{model}.
We modeled a set of parameter configurations to determine which scenario is the most plausible for interpreting our observations. Although this is an idealized experiment, where radiative transfer is simplified and chemistry is parametrized, it provides a useful framework for understanding our observations. It suggests that to explain the observed velocity difference between \n2dp(2-1) and \pnh2d, the interpretation based on the drift velocity in the presence of asymmetric infall/contraction cannot be excluded, as well as the interpretation where the observations are produced by a combined effect of drift velocity and different abundance distributions of the two molecules.
The detailed models, analyses, and results of this experiment are presented in \citet{Grassi2026}.

\section{Summary and conclusions}\label{conc} 

The analyses of the velocity structure at $\sim0.1$\,pc scale towards the L1544 prestellar core from the deuterated  \n2dp(2-1) ion and \pnh2d$(1_{11}-1_{01})$ neutral molecules show a mean difference of $0.05\,$\,km/s  with a standard deviation  of $0.02\,$\,km/s (Figure\,\ref{Fig:Vdrift}). 
This observed velocity difference may be tracing the decoupling between ions  and neutrals within the core hinting at the ongoing ambipolar diffusion and the onset of the core collapse.
The observed velocity difference is supported by theoretical calculations of evolution of the ambipolar diffusion coefficient in dense core environments including  dust grain growth (cf. Section\,\ref{diss1}). 
{\revn In addition to the difference of the centroid velocity,  the difference between the linewidths of ions and neutrals is an observational signature of the presence of ion-neutral drift and ongoing ambipolar diffusion. 
In our observations, we do not detect a significant difference between the linewidths for \n2dp and \pnh2d. This may be attributed to the subsonic infall motions  of the gas in L1544, that would mask the expected broadening of the linewidth of the neutrals, which are infalling faster compared to the ions better coupled with the B-field.  The absence of linewidth differences may also be due to geometric effects in the presence of inclination of the system with respect to the POS. 
}

Although recent calculations from \citet{Fukihara2026}
 support the interpretation of our observational results as tracing ambipolar diffusion, we cannot completely exclude  the possibility that the two species we have studied may be tracing slightly different regions and densities, and thus the dynamical properties of different regions within the core.
To better describe the physical and chemical properties of L1544, high-angular resolution data with interferometers (ALMA/NOEMA) and the next generation large single dish telescopes (AtLAST) equipped with high-spectral resolution (10\,kHz) spectrometers are needed. Future  observations  of ion-neutral pairs towards  a large sample of prestellar cores 
at different evolutionary stages and in different environments, 
compared to  simulations with ambipolar diffusion will help  better constrain the origin of the observational results.  
{\revn Such statistical studies will also provide the data needed to assess the impact of the magnetic field geometry and density structure, as well as  the effect of projections/inclinations on the observed velocities and linewidths derived from LOS integration. }

Finally, we propose that measurements of ion-neutral drift velocities within dense filaments and cores, may provide new constraints on the 3D magnetic field strength and the evolution of dust size distribution within prestellar cores on their way to star formation. 

\begin{acknowledgements}
S.S., T.G., P.C., J.P., S.J., and A.I. gratefully acknowledge the support of the Max Planck Society. 
This work was supported by the NINS-DAAD International Personal Exchange Program (2023-2025). This program is a joint funding initiative of the National Institutes of Natural Sciences (NINS) in Japan and the German Academic Exchange Service (DAAD).
\end{acknowledgements}

\bibliographystyle{aa}
\bibliography{bibfile}

\begin{appendix}

\onecolumn
\section{Fitting results}\label{App1}

Figure\,\ref{Fig:SpectraFit} shows the  \n2dp$(2-1)$  and \pnh2d\  spectra towards the  central pixel of the maps (RA$_{J2000.0}= 05^h04^m16.539^s$ and  Dec$_{J2000.0}=25^\circ10'47\parcs54$). The red profile shows the best fit model. The velocity  resolutions are 0.0379\,km/s for \n2dp$(2-1)$  and 0.0266\,km/s for  \pnh2d. The rest   (unsplitted) frequencies are 154217.0112\,MHz %
for \n2dp   and 110153.5940\,MHz %
for  \pnh2d.  The $V_{\rm lsr}$ for both cubes was set to 7.2\,km/s. All the spectra show a single velocity component.

\begin{figure*}[ht!]
   \centering
  \resizebox{8.9cm}{!}{\includegraphics[angle=0]{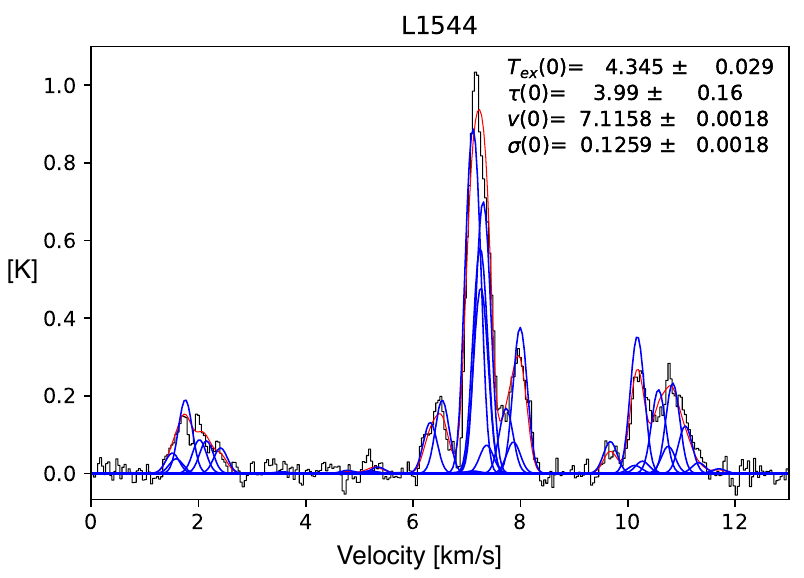}}
\resizebox{8.9cm}{!}{\includegraphics[angle=0]{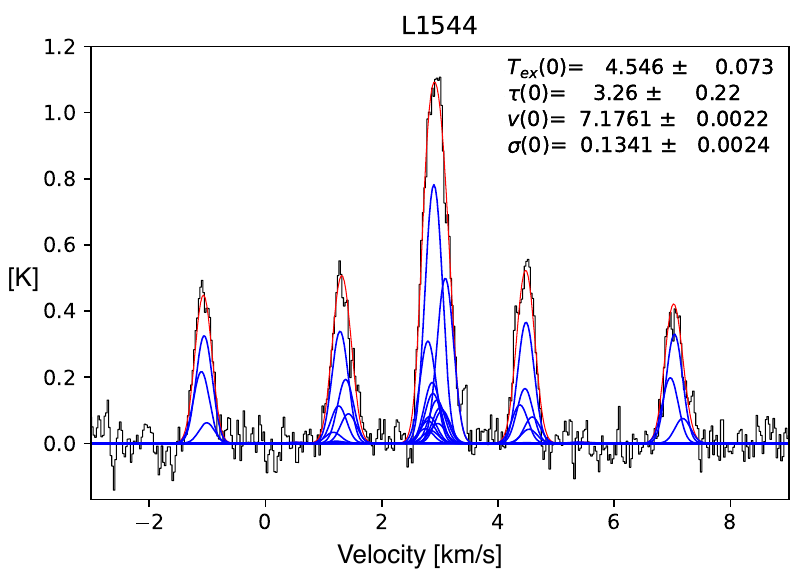}}
  \caption{Observed \n2dp$(2-1)$ ({\it left}) and \pnh2d ({\it right}) spectra in black towards the central position of the map. The best fit obtained with $pyspeckit$ is shown in red. The blue profiles are the best  fits for all the hyperfines. The results of the fit are indicated in the top right corner of each plot, with $T_{\rm ex}$ the excitation temperature in K,  $\tau$
  the optical depth, $V$ the velocity in km/s, and $\sigma_{\rm v,obs}$ the velocity dispersion in km/s. 
  }          
  \label{Fig:SpectraFit}
\end{figure*}

Figure\,\ref{Fig:histo_V_Err} shows the distributions of the $1\sigma$ uncertainty on the centroid velocities and on the uncertainty on the velocity difference  derived from the $pyspeckit$ fit (cf. Section\,\ref{SpecFit}). The main uncertainties on the order of $\sim0.006$\,km/s are an order of magnitude smaller than the observed velocity difference (cf. Section\,\ref{results}).

\begin{figure*}[ht!]
   \centering
  \resizebox{8.3cm}{!}{\includegraphics[angle=0]{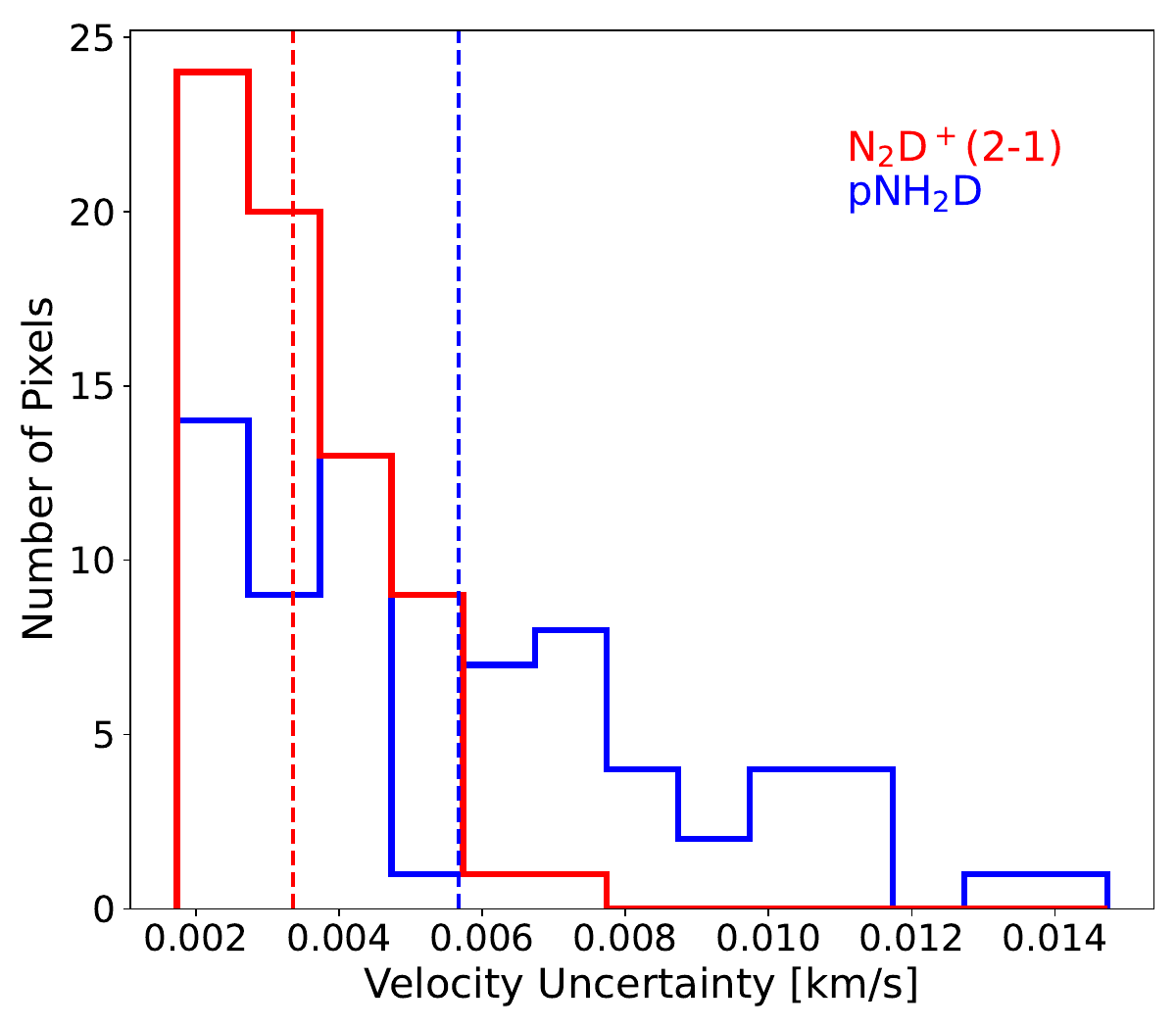}}
    \resizebox{8.3cm}{!}{\includegraphics[angle=0]{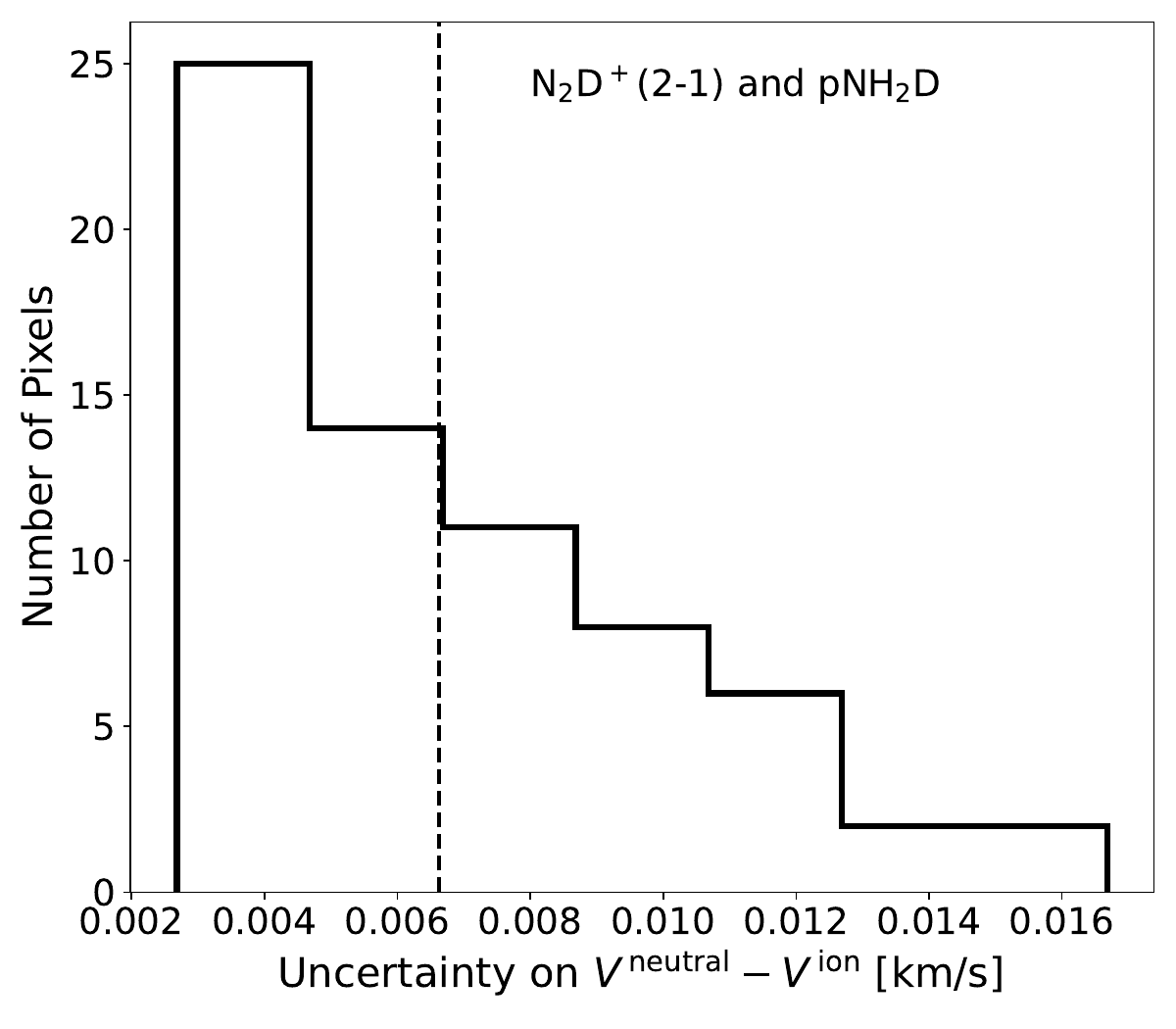}}
  \caption{
   {\it Left:}  Distribution of the $1\sigma$ uncertainty on the centroid velocity \n2dp and \pnh2d derived from the $pyspeckit$ fit.
     The mean values of these uncertainties, indicated with the vertical lines, are 0.0033 and 0.0056\,km/s for the \n2dp\ and \pnh2d\ data points, respectively.
     {\it Right:} Distribution of uncertainty on $V^{\rm neutral}-V^{\rm ion}$ derived from error propagation. The mean value of the distribution is 0.006\,km/s.
  }          
  \label{Fig:histo_V_Err}
\end{figure*}

Figure\,\ref{Fig:TexMaps} shows  {\rev the integrated intensity maps} and  the excitation temperature  $T_{\rm ex}$ maps of  \n2dp\ and \pnh2d, {\rev displaying the highly centrally concentrated emission of the ion-neutral pair}.  {\revn The integrated intensity radial profiles of \n2dp\ and \pnh2d\ (Figure\,\ref{Fig:Radprof}) are very similar, with only a $\sim20\%$ larger value for \pnh2d\ within the central part of the core (radii\,$<0.01$\,pc).}
  {\revn On the $T_{\rm ex}$ maps, 
\n2dp\ appears  slightly} ($\sim25\%$) more extended  compared to that of \pnh2d. The mean and standard deviation of the $T_{\rm ex}^{\rm neutral}/T_{\rm ex}^{\rm ion}$ ratio and the $\tau^{\rm neutral}/\tau^{\rm ion}$ are $1.03\pm0.44$ and  $0.99\pm0.18$ (Figure\,\ref{Fig:Tex_tau_ratio})  further supporting that these two molecules are tracing similar optical depth and  densities.

\begin{figure*}[ht!]
   \centering
     \resizebox{8.3cm}{!}{\includegraphics[angle=0]{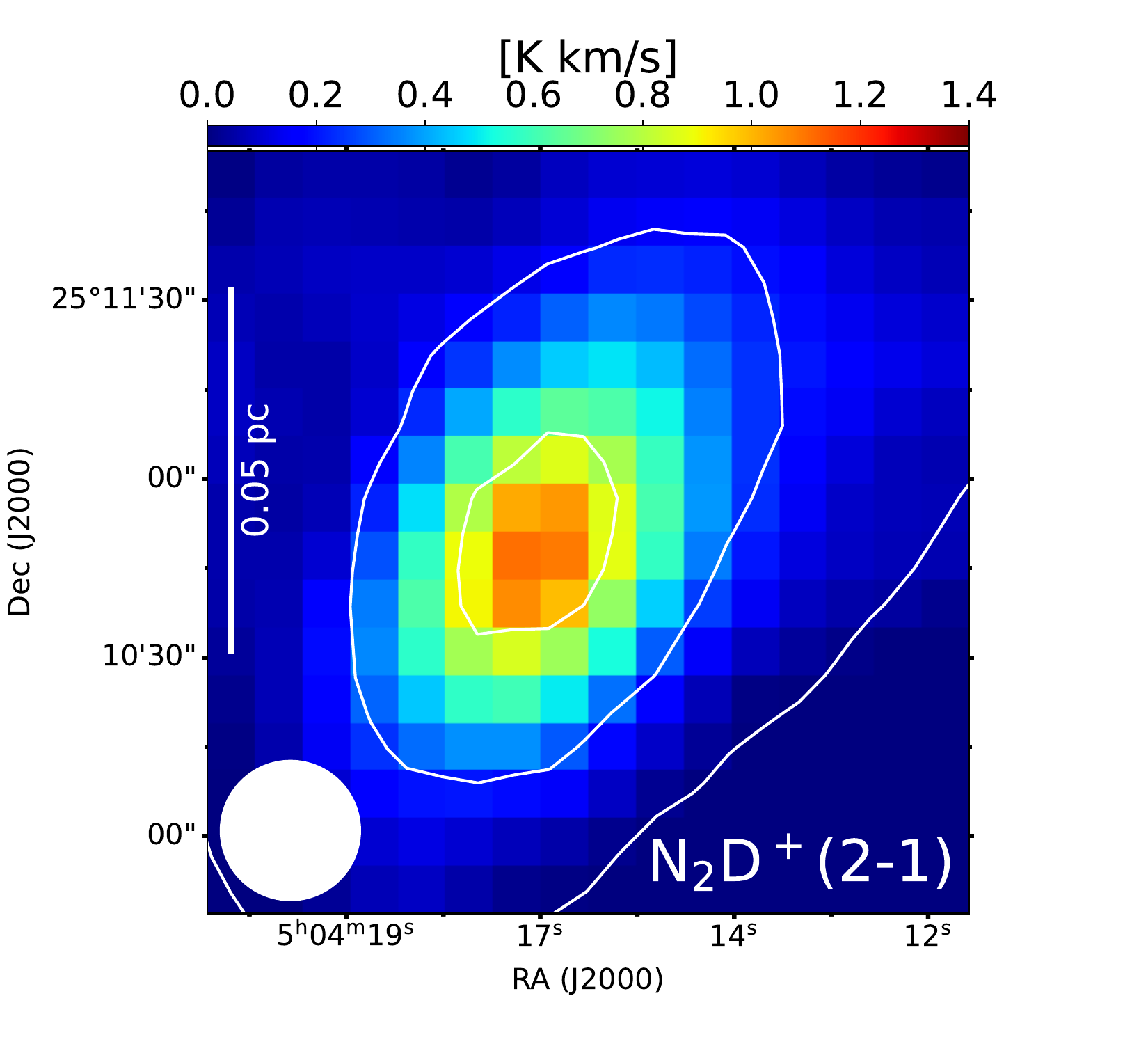}}
\resizebox{8.3cm}{!}{\includegraphics[angle=0]{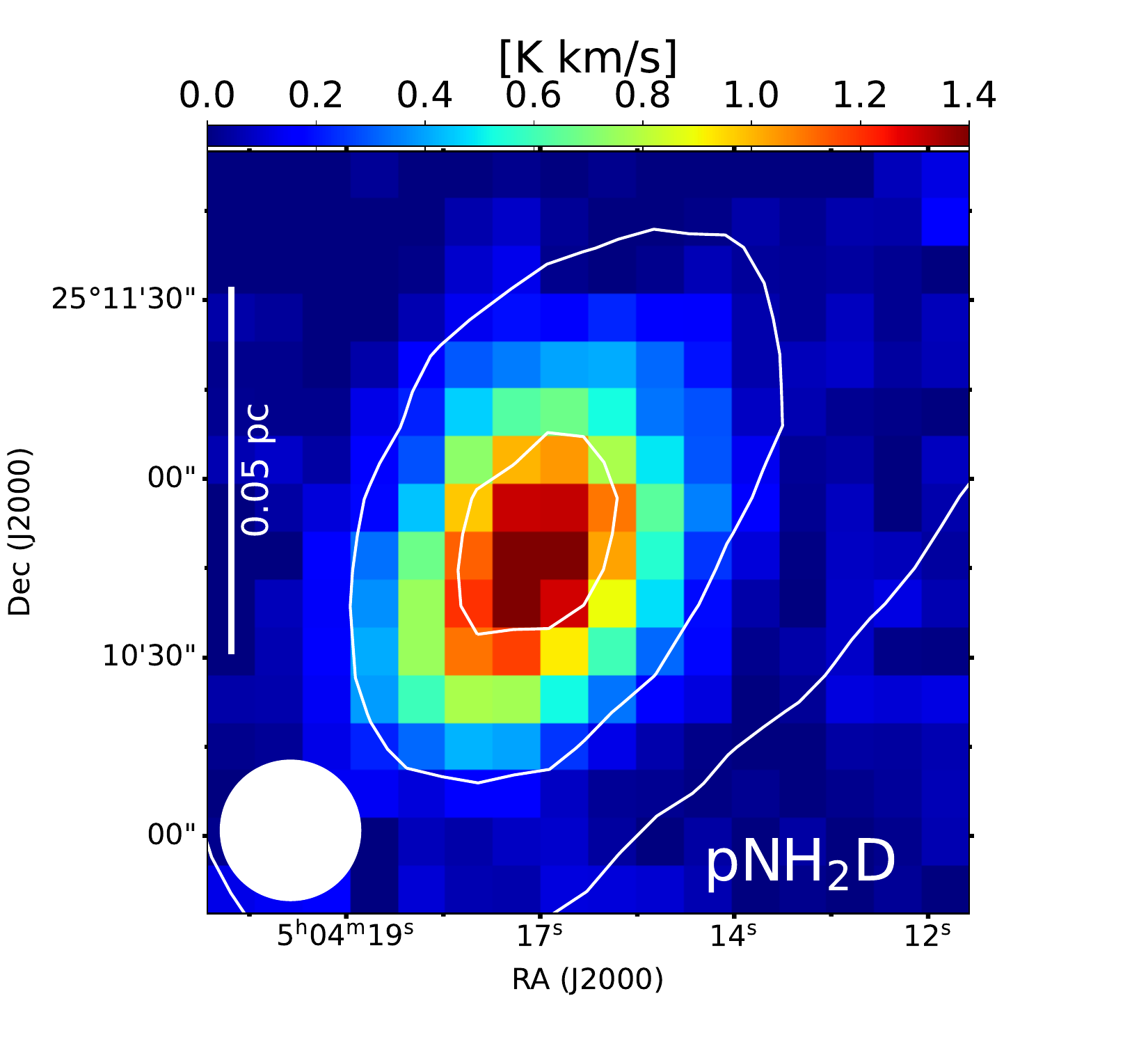}}
  \resizebox{8.3cm}{!}{\includegraphics[angle=0]{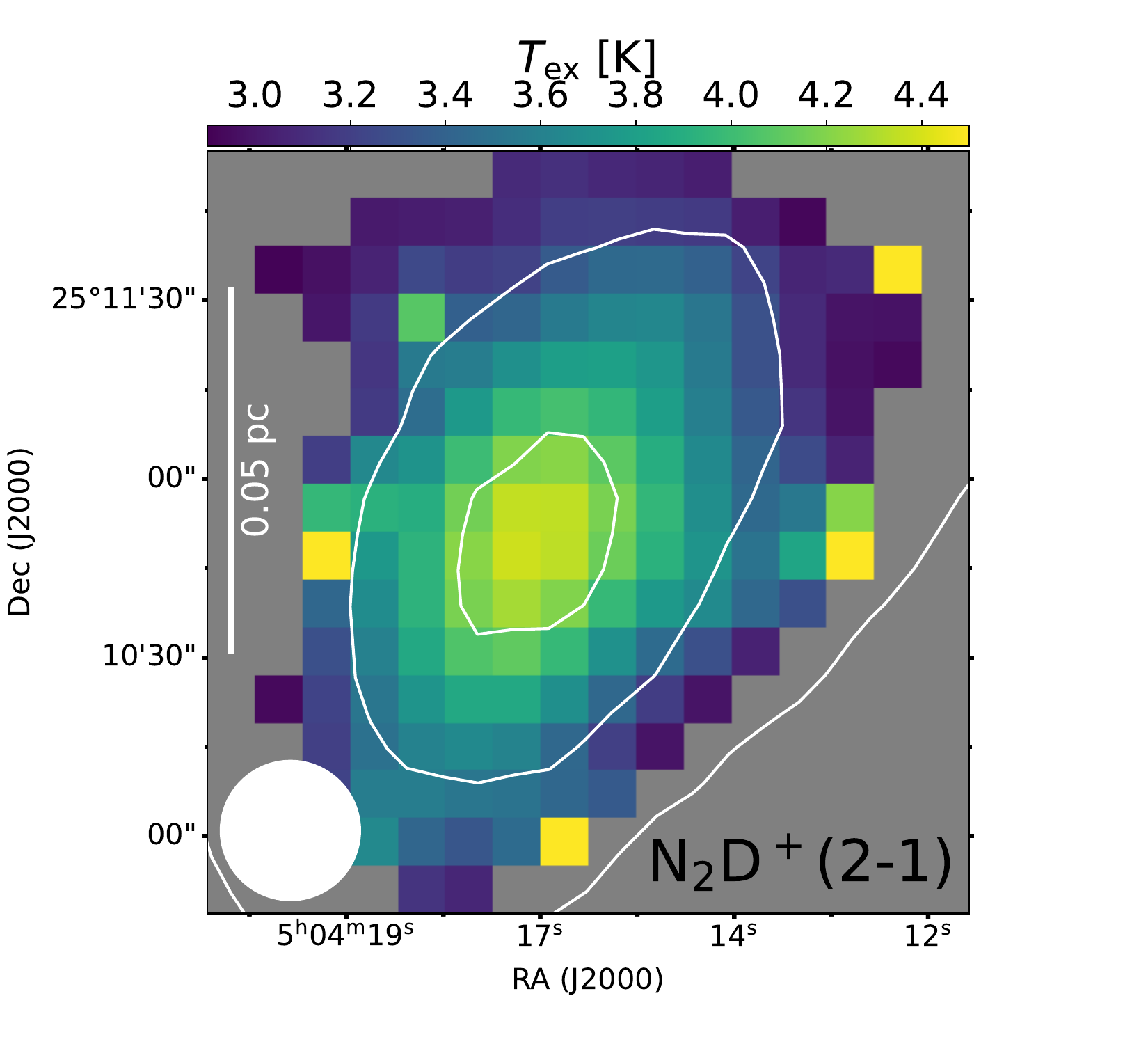}}
\resizebox{8.3cm}{!}{\includegraphics[angle=0]{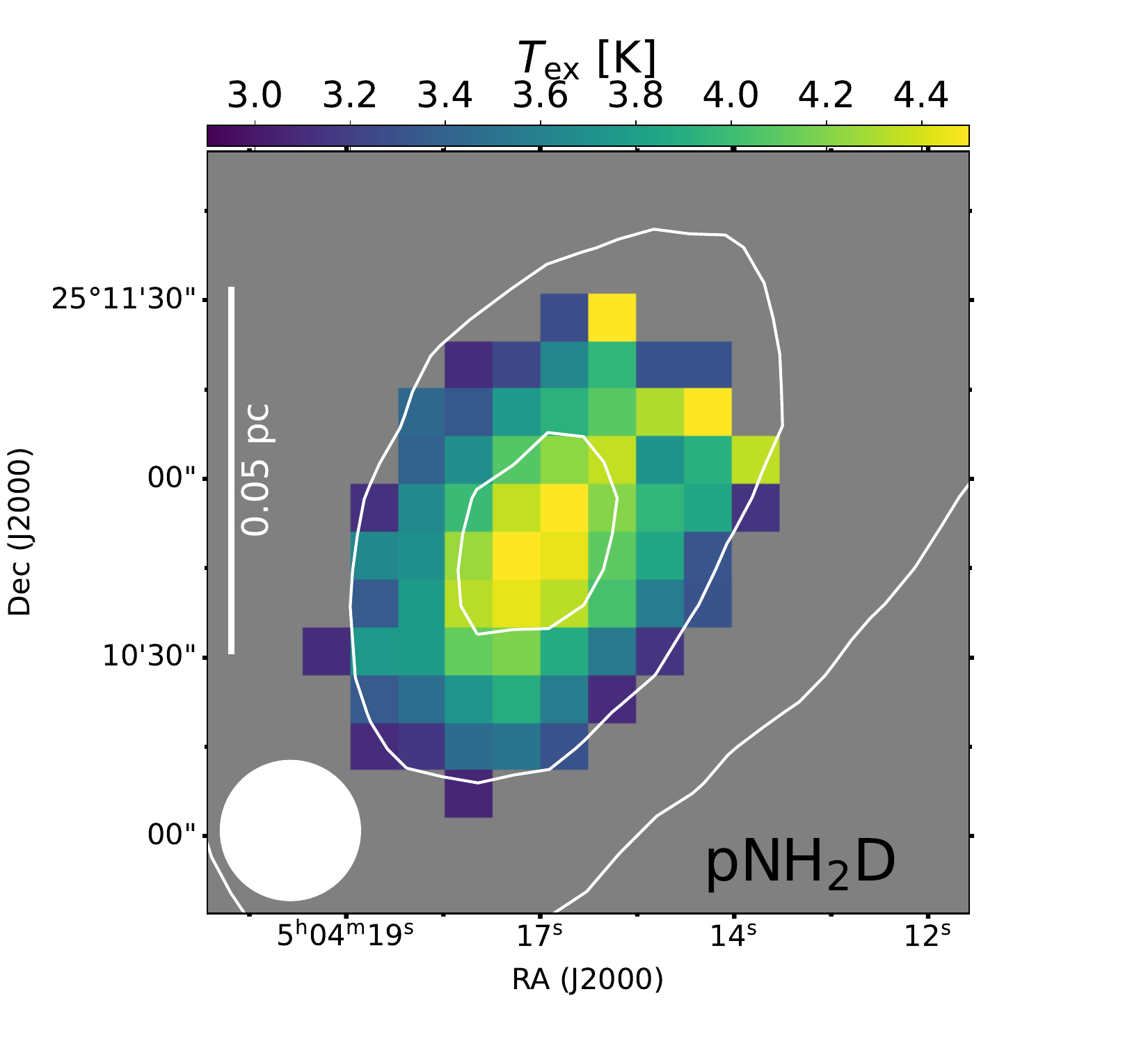}}
  \caption{ {\rev {\it Top:}  Integrated intensity maps for \n2dp\ ({\it left}) and \pnh2d\ ({\it right}).  {\it Bottom:} }
 Excitation temperature ($T_{\rm ex}$) maps derived from pixel-per-pixel fitting of the  \n2dp\ ({\it left}) and \pnh2d\ ({\it right}) spectra with $pyspeckit$ (cf. Section\,\ref{SpecFit}). The white contours at $0.8, 1.7,$\,and $2.5\times10^{22}$\,cm$^{-2}$  trace the gas column density  as derived from \herschel\ data \citep[from][]{Spezzano2017} and are the same on all panels.   The white filled circles show the 23\parcs5  beam size (0.019\,pc at the 170\,pc distance of L1544) 
  {\rev and the white vertical lines indicate the 0.05\,pc scale.} 
  }          
  \label{Fig:TexMaps}
\end{figure*}

\begin{figure*}[ht!]
   \centering
     \resizebox{8.3cm}{!}{\includegraphics[angle=0]{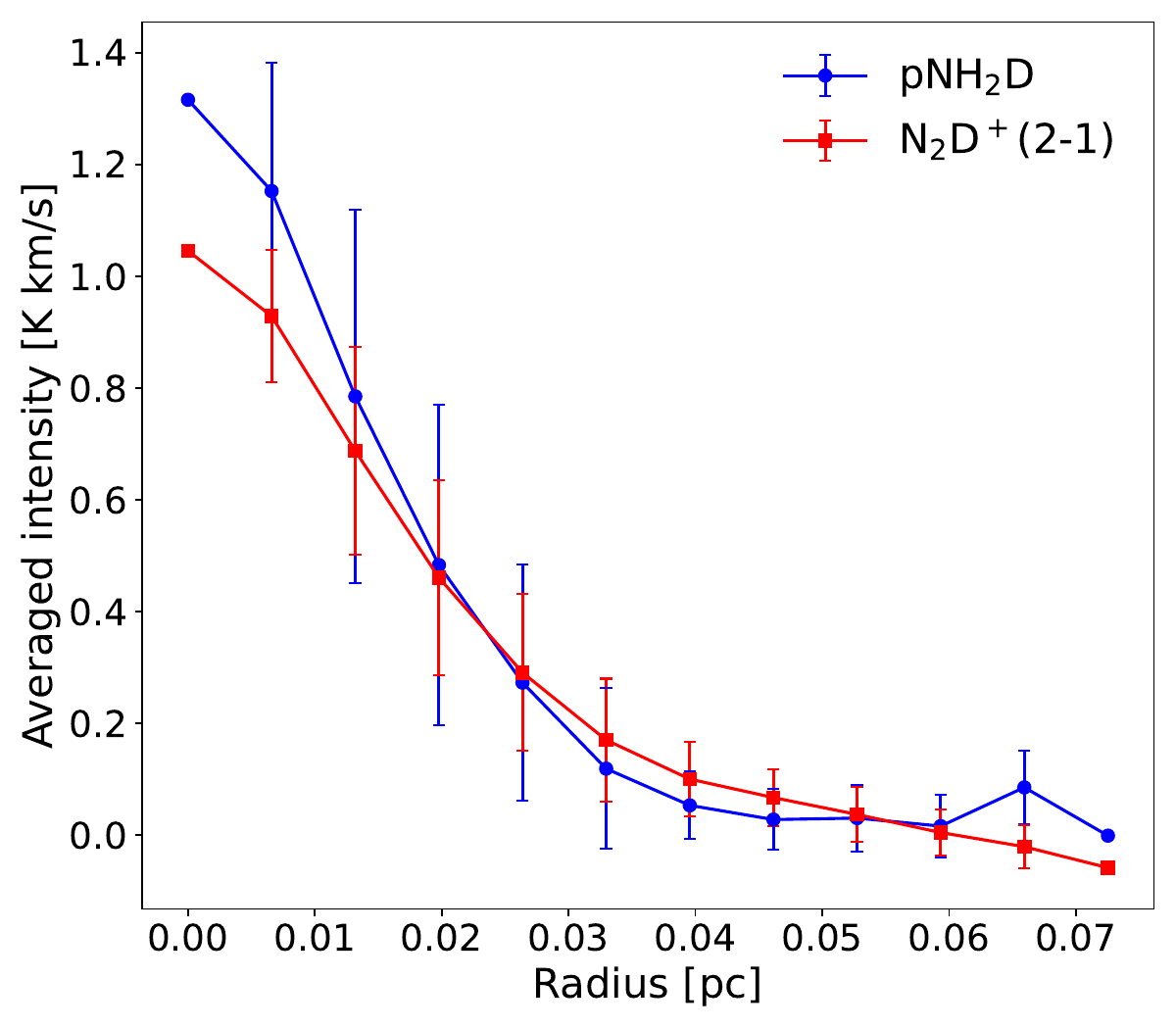}}
  \caption{  {\revn Integrated intensity radial profile of \n2dp\ (red) and \pnh2d\ (blue). The profiles are obtained by circularly averaging the radial profiles about the 
  central pixel of the integrated intensity maps (cf.\,Figure\,\ref{Fig:TexMaps}). The vertical bars indicate the standard deviation on the azimuthal average at each radial distance.    
  }
  }          
  \label{Fig:Radprof}
\end{figure*}

\begin{figure*}[ht!]
   \centering
  \resizebox{8.3cm}{!}{\includegraphics[angle=0]{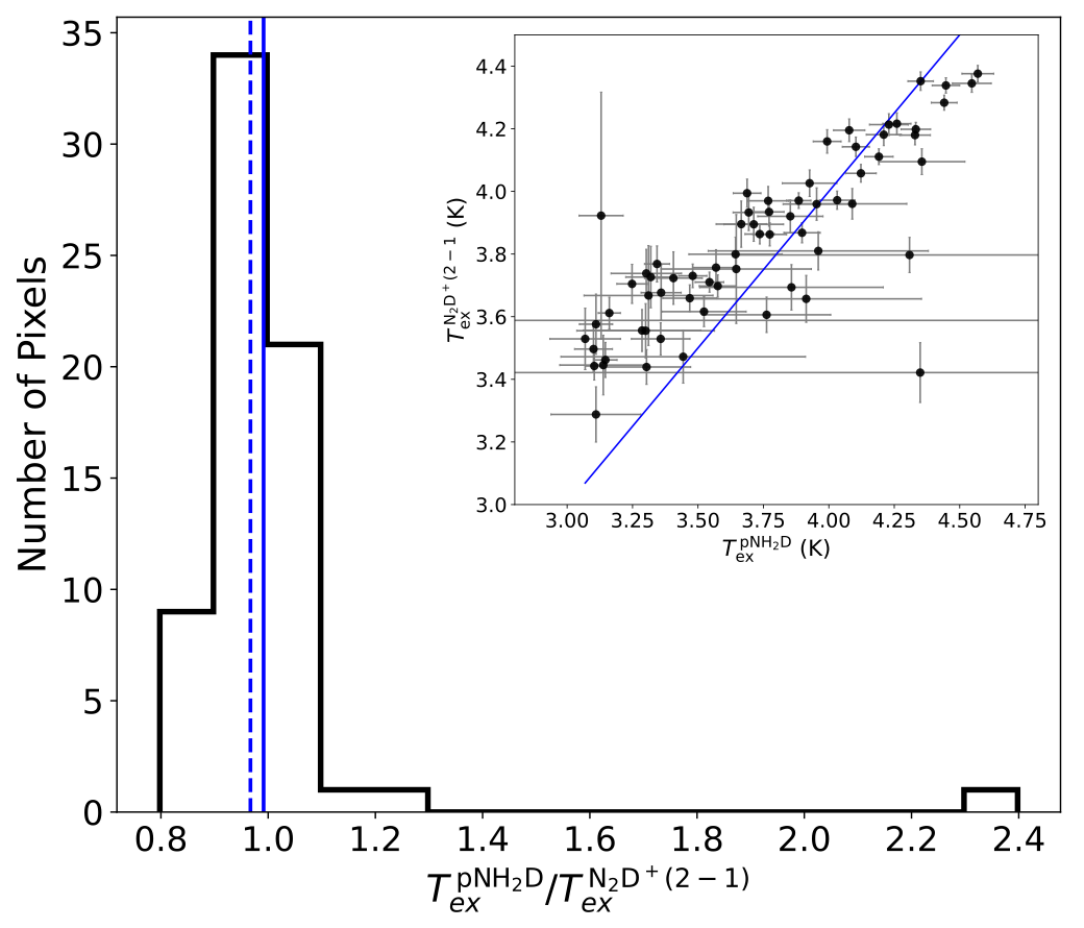}}
    \resizebox{8.3cm}{!}{\includegraphics[angle=0]{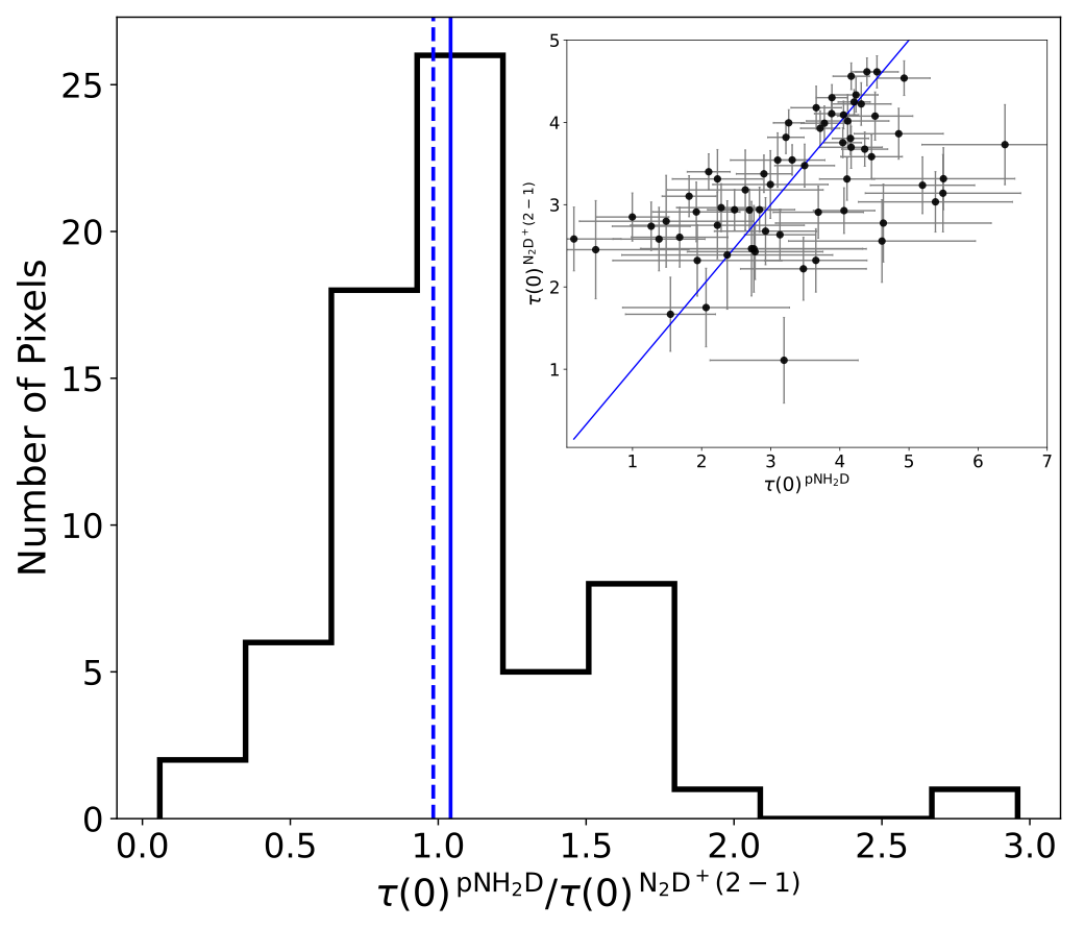}}
  \caption{
   {\it left:}  Histogram of the $T_{\rm ex}$ ratio  of \pnh2d and \n2dp(2-1). The mean, median, and standard deviation of the distribution are  0.99, 0.97, and 0.18\,K, respectively.  The inset shows the scatter plot of the $T_{\rm ex}$ values and their uncertainties. The oblique line indicates the 1:1 relation.  {\it Right:}  Similar to the left panel for the $\tau(0)$ values. The mean, median, and standard deviation  of the distribution  are 1.03, 0.98, 0.44, respectively. 
  }          
  \label{Fig:Tex_tau_ratio}
\end{figure*}

\newpage
\section{Total velocity dispersion}\label{Sect:totVel}

To compare the velocity dispersions of the two molecules of interest \n2dp\ and \pnh2d, we estimate the total velocity dispersion of the mean free particle with a molecular weight of $\mu_{\rm p}=2.33$ in molecular clouds \citep{Kauffmann2008}.  
We derive the  non-thermal velocity dispersion of the gas by subtracting the thermal velocity dispersion from the linewidth measured for each species, assuming that the two contributions are independent of each other and could be added in quadrature \citep{Myers1983}. The thermal velocity dispersion of each species is given by
 $ \sigma_\mathrm{T}(\mu_{\rm obs})=\sqrt{k_{\rm B}T/\mu_{\rm obs}m_{\rm H}}, $ 
where $k_{\rm B}$ is the Boltzmann constant, $m_{\rm H}$ is the hydrogen mass, $\mu_{\rm obs}$ is the atomic weight of the observed molecule with $\mu_{\rm obs}=30$ for \n2dp\ and $\mu_{\rm obs}=18$ for \pnh2d. We adopted a gas temperature of 10\,K which is the mean gas temperature of L1544 at the observed scales and densities \citep{Crapsi2007,Spezzano2017}. 
The non-thermal velocity dispersion is calculated as 
$\sigma_{\rm v,NT}=\sqrt{\sigma^2_{\rm v,obs}-\sigma^2_\mathrm{T}(\mu_{\rm obs})},$
where $\sigma_\mathrm{T}$ is the thermal velocity dispersion of the observed species with $\sigma_\mathrm{T}\sim 0.053$ and $0.068$\,km/s for \n2dp\ and \pnh2d, respectively 
and $\sigma_{\rm obs}$ is the observed velocity dispersion for each molecule derived from the fit (Section\,\ref{SpecFit}).  
The total velocity dispersion of the mean free particle ($\mu_{\rm p}=2.33$) and thermal velocity dispersion $\sigma_\mathrm{T}=\sqrt{k_{\rm B}T/\mu_{\rm p} m_{\rm H}}\sim0.19\pm0.02$\,km/s for $T=10\pm2$\,K is given by
$ \sigma_{\rm v,tot}=\sqrt{\sigma^2_{\rm v,NT}+\sigma^2_\mathrm{T}}.$
We derive maps of  $ \sigma_{\rm v,tot}$ and  $\sigma_{\rm v,NT}$ for both \n2dp\ and \pnh2d. 
{\rev Figure\,\ref{Fig:S_tot_NT} shows the distributions of  $ \sigma_{\rm v,tot}$ and  $\sigma_{\rm v,NT}$.  
Figure\,\ref{Fig:Sdiff} shows the map of the  difference of the ion-neutral  total velocity dispersion  ($\delta \sigma_{\rm v,tot}=\sigma_{\rm v,tot}^{\rm neutral}-\sigma_{\rm v,tot}^{\rm ion}$) and the    histograms of $\delta \sigma_{\rm v,tot}$  and $\delta \sigma_{\rm v,NT}$. 
}

\begin{figure*}[ht!]
   \centering
  \resizebox{8.9cm}{!}{\includegraphics[angle=0]{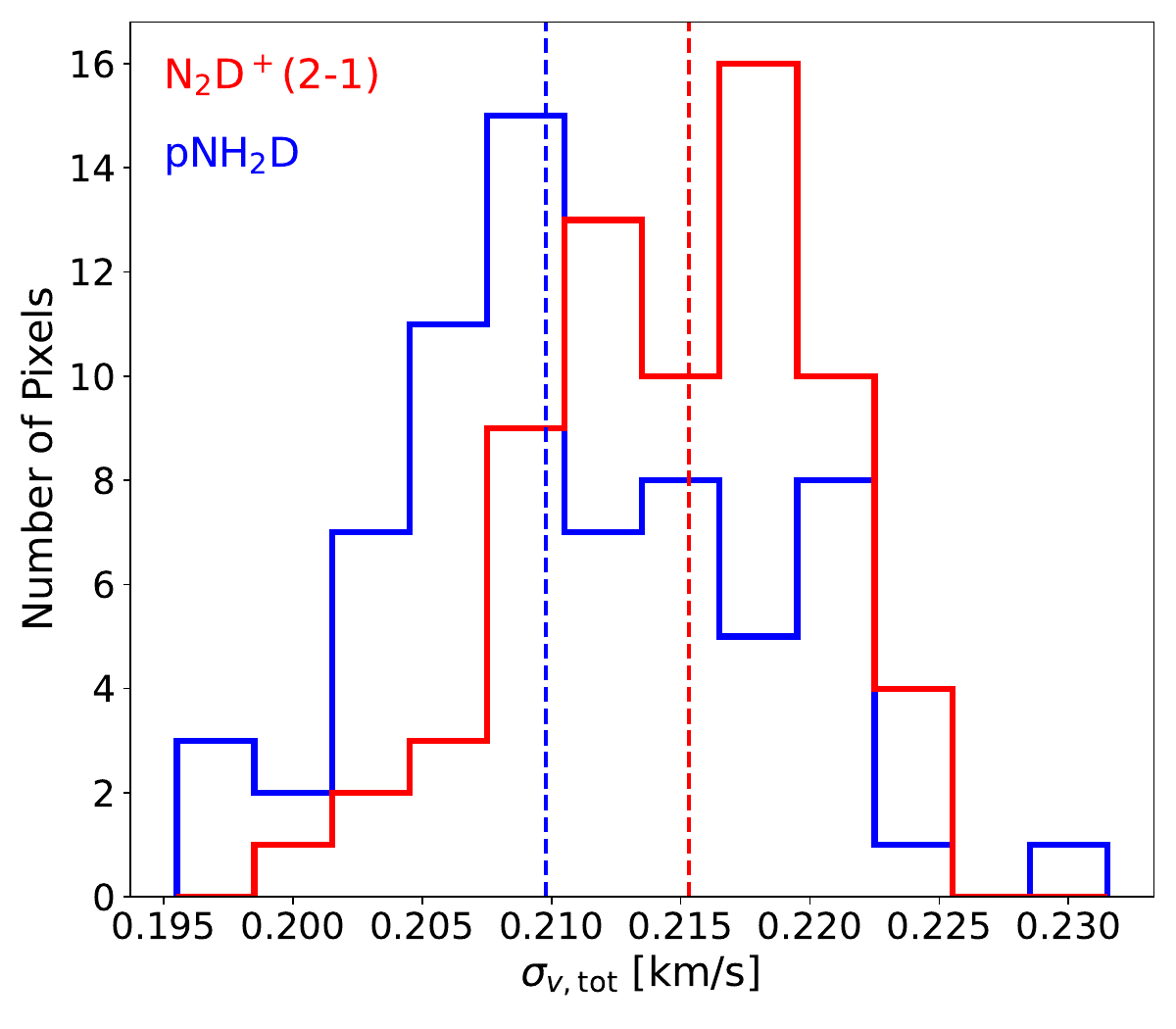}}
\resizebox{8.9cm}{!}{\includegraphics[angle=0]{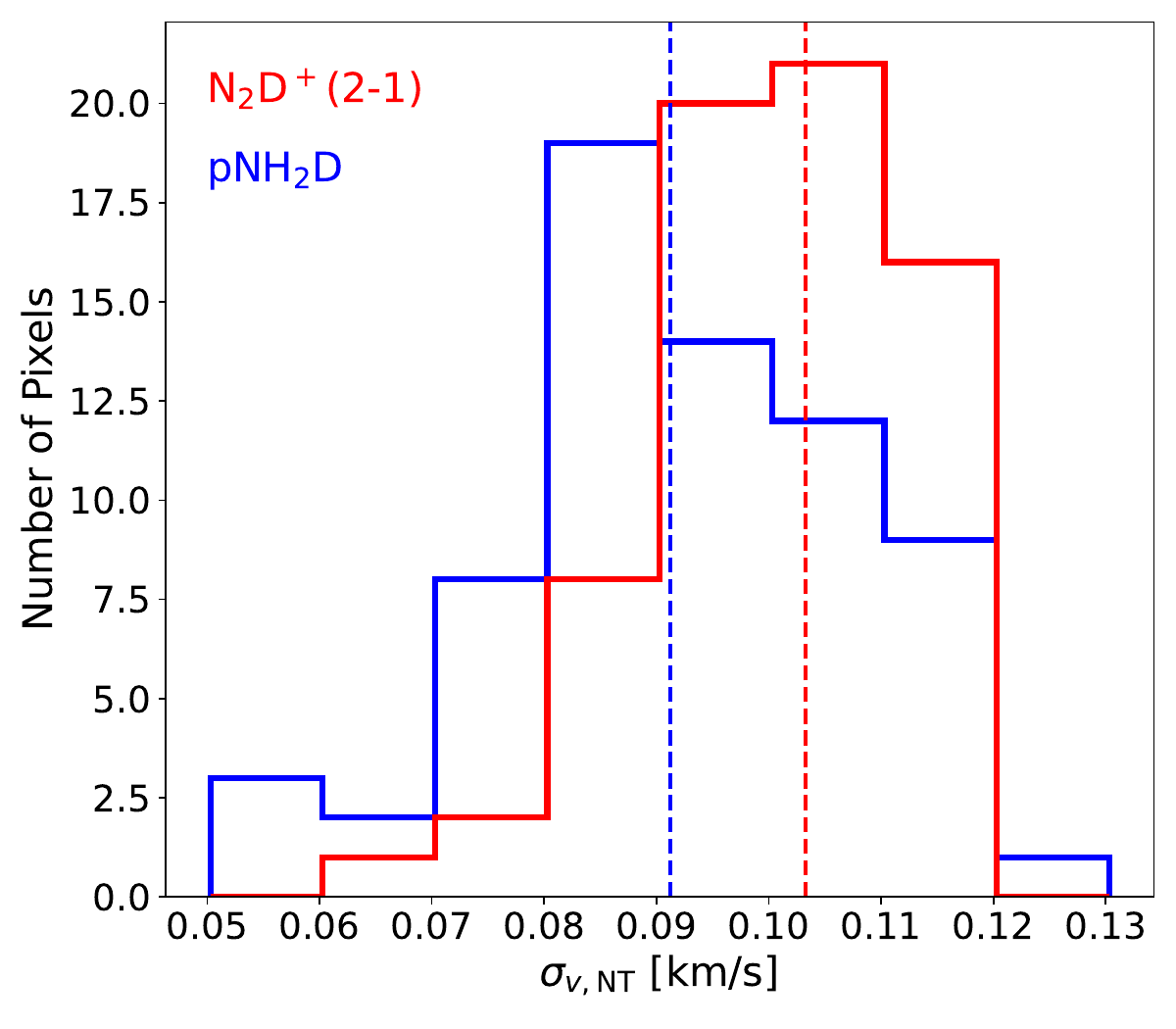}}
  \caption{{ \rev
   {\it Left:}
  Histograms of $\sigma_{\rm v,tot}$ for \n2dp in red and \pnh2d in blue. 
  The median and standard deviation of the  distributions are 0.215\,km/s and 0.005\,km/s for \n2dp and 0.210\,km/s and 0.007\,km/s for \pnh2d. 
    {\it Right:}
  Histograms of $\sigma_{\rm v,NT}$ for \n2dp in red and \pnh2d in blue. 
  The median and standard deviation of the  distributions are 0.103\,km/s and 0.012\,km/s for \n2dp and 0.091\,km/s and 0.016\,km/s for \pnh2d.
    }       }   
  \label{Fig:S_tot_NT}
\end{figure*}

\begin{figure*}[ht!]
   \centering
  \resizebox{8.9cm}{!}{\includegraphics[angle=0]{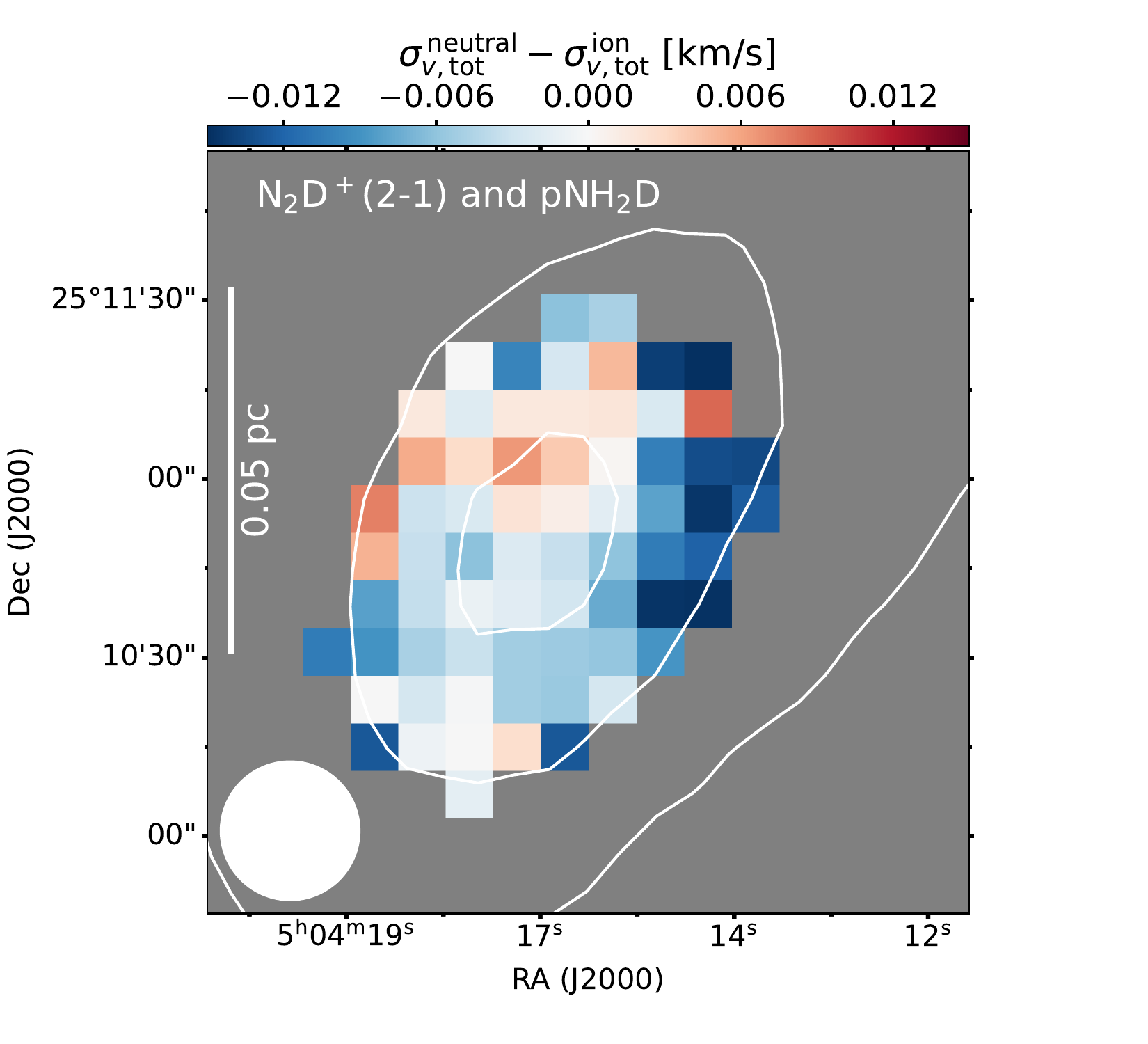}}
\resizebox{8.9cm}{!}{\includegraphics[angle=0]{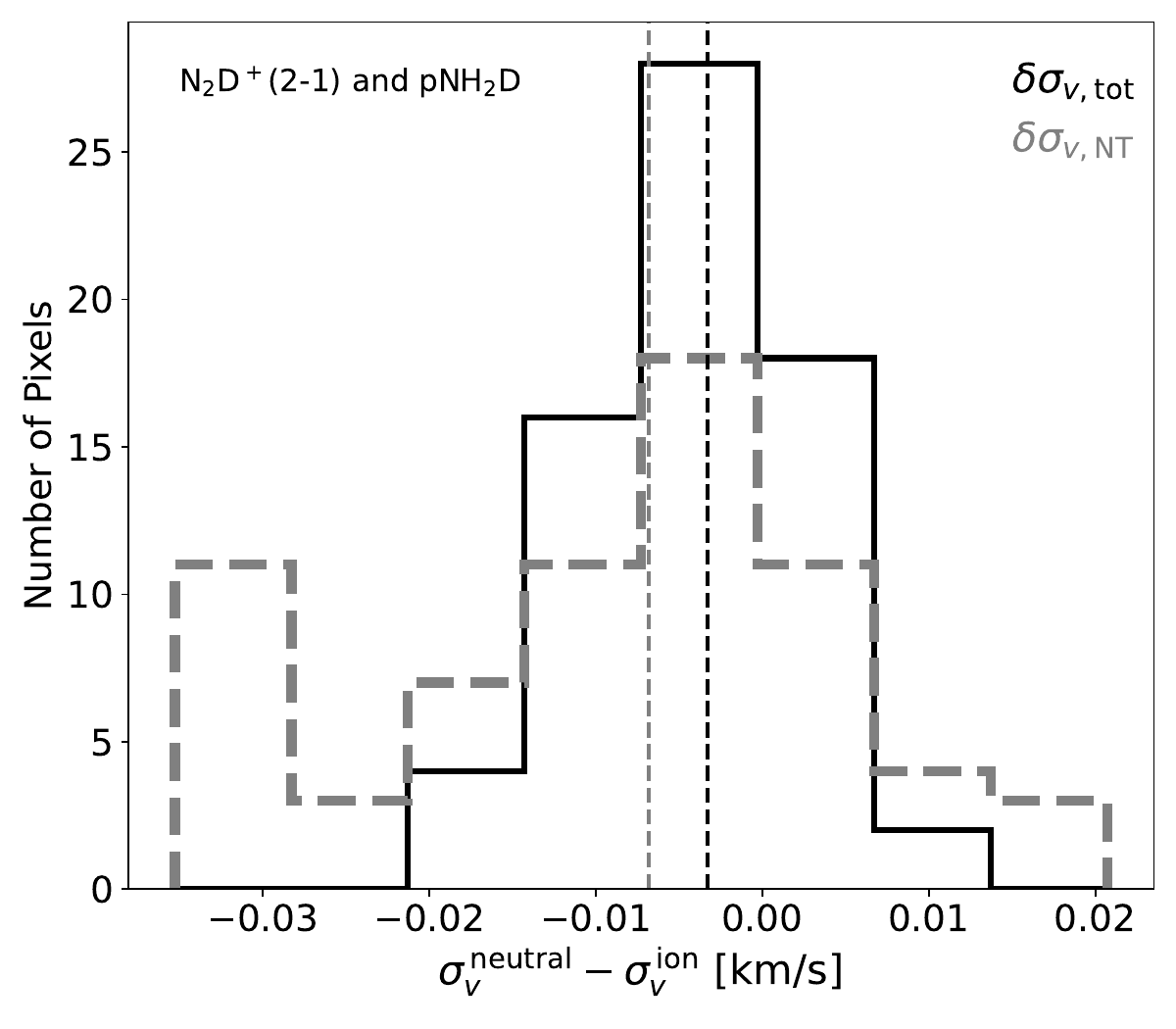}}
  \caption{
 {\it Left:} Map of the  difference between the  neutral (\pnh2d) and the ion  (\n2dp)  total velocity dispersion  ($\delta \sigma_{\rm v,tot}=\sigma_{\rm v,tot}^{\rm neutral}-\sigma_{\rm v,tot}^{\rm ion}$).   {\it Right:} Histogram of $\delta \sigma_{\rm v,tot}$ in black and of $\delta \sigma_{\rm v,NT}$ in gray. 
  {\rev The vertical dashed black and gray lines indicate the mean values for the $\delta \sigma_{\rm v,tot}$ and  $\delta \sigma_{\rm v,NT}$ distributions, respectively. 
 }
 The mean and standard deviation  for   $\delta \sigma_{\rm v,tot}$  are $-0.004$\,km/s and $0.006$\,km/s, and for   $\delta \sigma_{\rm v,NT}$  are $-0.007$\,km/s and $0.014$\,km/s, respectively.
  }          
  \label{Fig:Sdiff}
\end{figure*}

\section{Comparing with N$_2$D$^+(1-0)$}\label{App_comp}

We here compare the results of the $pyspeckit$ fits of the \pnh2d, \n2dp$(2-1)$,  and \n2dp$(1-0)$ PPV cubes. 
The rest frequency of \n2dp$(1-0)$ is 77.1096162\,MHz and the spectral resolution of the cube is 0.076\,km/s. 
 All the cubes have been smoothed to the coarser 34"  angular  resolution of \n2dp$(1-0)$. 
 We do not resample the data spectrally and correct for the channel response to derive 
$\sigma_{\rm v}=\sqrt{\sigma_{\rm v,obs}^2-(\Delta_{\rm channel}/2.355)^2}$,  
where $\sigma_{\rm v,obs}$ is the velocity dispersion derived from the fit and $\Delta_{\rm channel}$ the  velocity resolution of the PPV cube. 
The critical densities computed for 10\,K  are $0.8\times10^5$, $1.4\times10^5$, and $4\times10^5$, for  \n2dp$(1-0)$, \pnh2d, and  \n2dp$(2-1)$, respectively.

Figure\,\ref{Fig:3mol_Tau_Tex} and Table\,\ref{table} compare the observed properties of \pnh2d, \n2dp$(2-1)$,  and \n2dp$(1-0)$.
The standard deviation of the distributions of 
 $T_{\rm ex}^{\rm neutral}/T_{\rm ex}^{\rm ion}$  and the $\tau^{\rm neutral}/\tau^{\rm ion}$ ratios 
 when the ion is \n2dp$(1-0)$  are broader  compared to those when the ion is \n2dp$(2-1)$, suggesting that the  \n2dp$(2-1)$ emission is more closely correlated to that of \pnh2d. The total velocity dispersion of \n2dp$(1-0)$  is  broader (by 0.013\,km/s) than that of \pnh2d, which may hint at the more turbulent gas and slightly lower density gas envelop that \n2dp$(1-0)$  may be tracing as opposed to that traced by \n2dp$(2-1)$ and \pnh2d.

\begin{figure*}[ht!]
   \centering
  \resizebox{8.9cm}{!}{\includegraphics[angle=0]{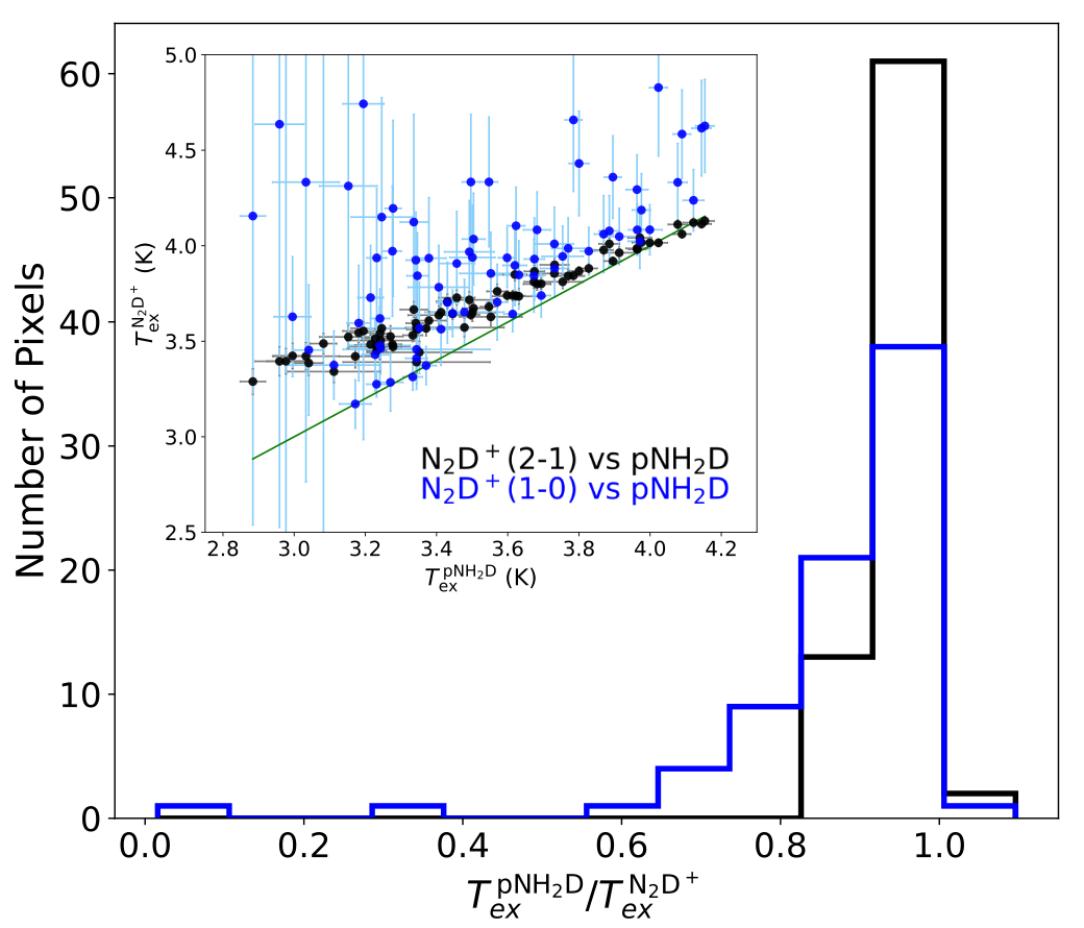}}
    \resizebox{8.9cm}{!}{\includegraphics[angle=0]{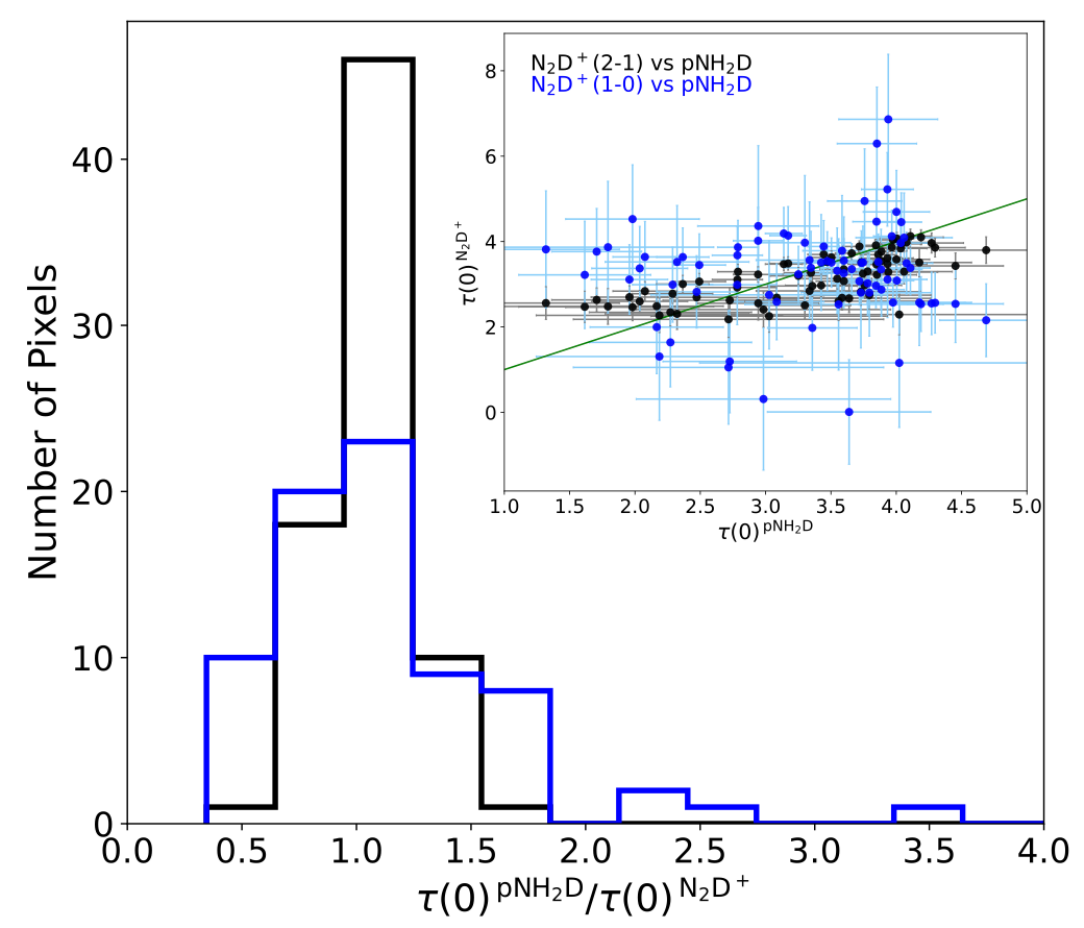}}
     \resizebox{8.9cm}{!}{\includegraphics[angle=0]{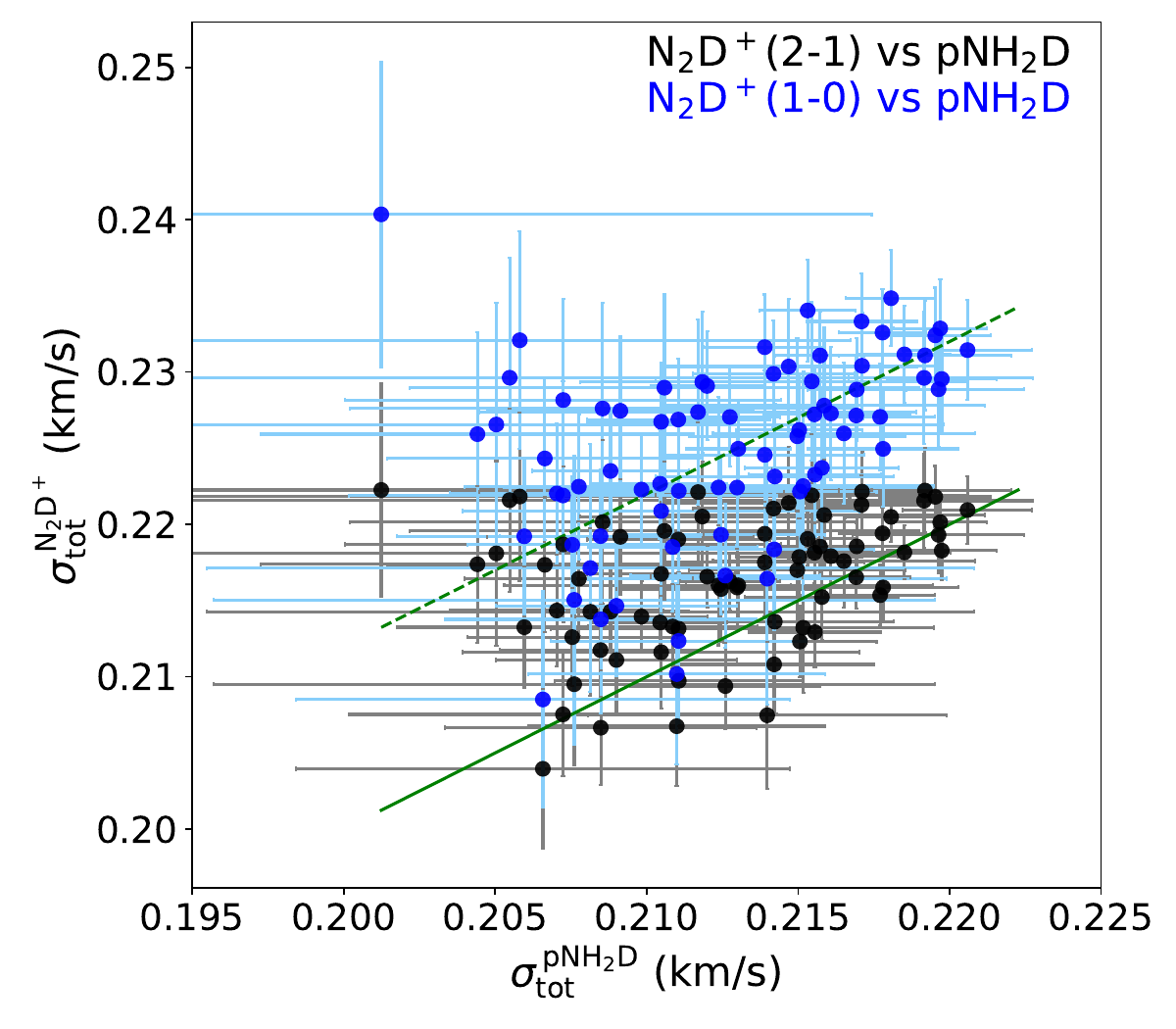}}
    \resizebox{8.9cm}{!}{\includegraphics[angle=0]{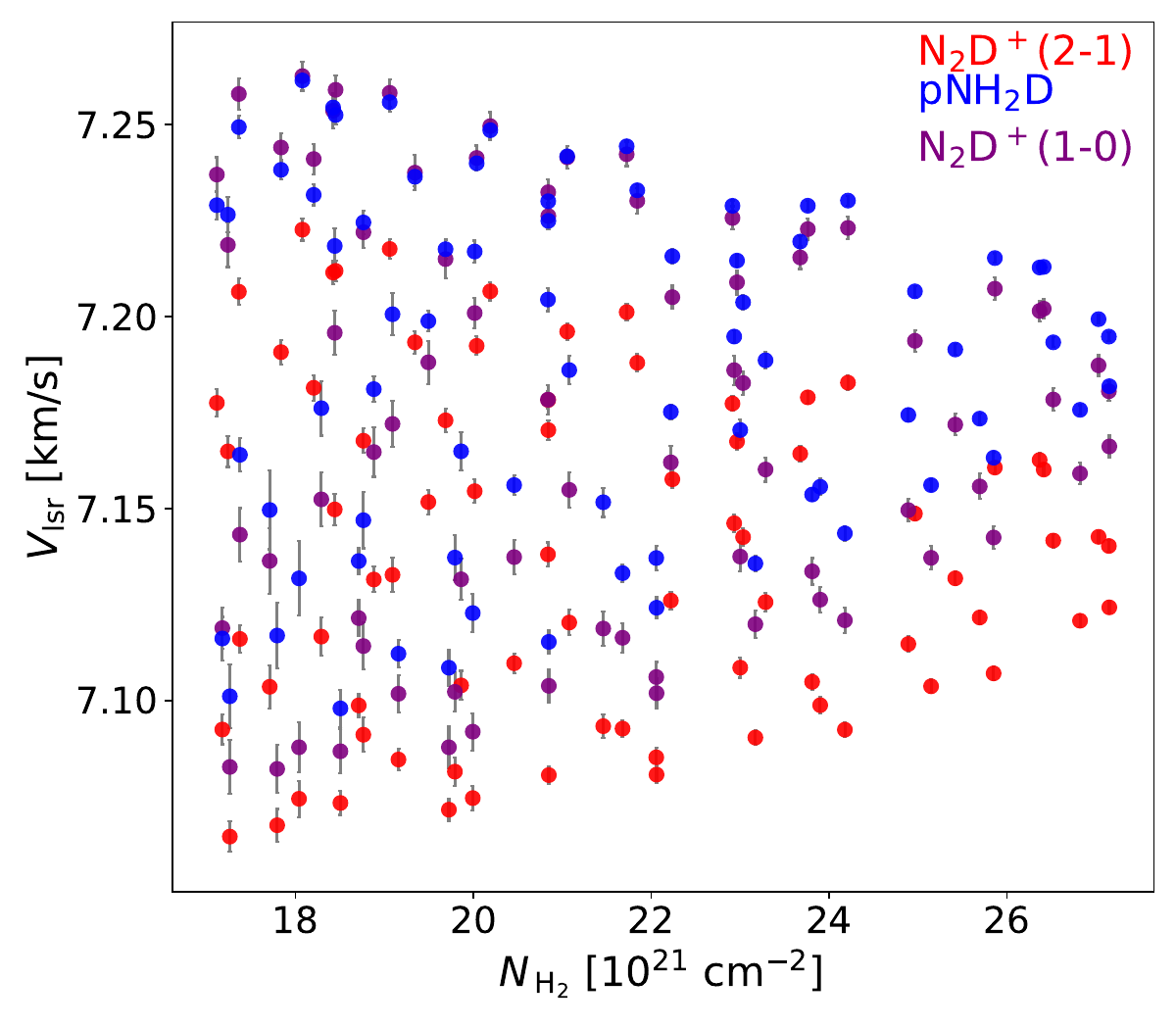}}
  \caption{
   {\it Top-left:}  Histrograms of the  $T_{\rm ex}^n/T_{\rm ex}^i$ ratio values, where the ion is \n2dp$(2-1)$  in black and \n2dp(1-0) in blue.  The velocity data cubes are all smoothed to the spatial resolution of \n2dp$(1-0)$   with HPBW of 34". The mean and standard deviation of the distributions are given in Table\,\ref{table}. The inset shows the  scatter plot of the $T_{\rm ex}$ values and their uncertainties.  The oblique  line indicates the 1:1 relation. 
   {\it Top-right:}  Similar  as the top-left panel  for the $\tau$ values.
   {\it Bottom-left:} Scatter plot of the velocity dispersion. The oblique solid and dashed blue lines indicate the 1:1 and the  $0.012$\,km/s relations, respectively. 
   {\it Bottom-right:} Scatter plot of  the  centroid velocity of \n2dp$(2-1)$  in red, \pnh2d in blue, and  \n2dp$(1-0)$  in purple   versus the $N_{\rm H_2}$ column density derived from \herschel\ %
   for data points with
   $N_{\rm H_2}>1.7\times10^{22}$\,cm$^{-2}$. %
   The vertical gray lines show the fitting uncertainties of $V_{\rm lsr}$. 
  }          
  \label{Fig:3mol_Tau_Tex}
\end{figure*}

\begin{table}
 \caption[]{ Table summarizing the mean and standard deviation of the distributions shown in Fig.\ref{Fig:3mol_Tau_Tex}.} 
    \label{table}
\tabcolsep 4pt
 {\centering \begin{tabular}{c|c|c|c|c}
 & $T^{\rm neutral}_{\rm ex}/T^{\rm ion}_{\rm ex}$  & $\tau^{\rm neutral}/\tau^{\rm ion}$ & $\sigma^{\rm neutral}_{\rm v,tot}-\sigma^{\rm ion}_{\rm v,tot}$ & $V^{\rm neutral}_{\rm lsr}-V^{\rm ion}_{\rm lsr}$\\
   &   &  & km/s & km/s  \\
   \hline
    \n2dp$(1-0)$ and \pnh2d& $0.88\pm0.14$ &$1.10\pm0.51$ &$-0.013\pm0.006$& $0.013\pm0.012$\\
    \n2dp$(2-1)$ and \pnh2d& $0.95\pm0.04$ &$1.05\pm0.20$ &$-0.004\pm0.005$&$0.050\pm0.010$ \\
  \end{tabular} \par}
\end{table}

\newpage
\section{3D differentiable analytical model of collapsing core}\label{model}

We have developed a 3D analytical model of a prestellar core controlled by a set of parameters. The versors $\mathbf{u}$ of the velocity vector field at each position $\mathbf{x} = (x, y, z)$ are determined by the $3\times3$ matrix $M$ (i.e., 9~parameters) as $\mathbf{u}=M\,\mathbf{x}$. For example, a symmetric free-fall is described by $M=-I$, where $I$ is the identity matrix. The velocity of the neutral molecule $\mathbf{v}$ is scaled from the versors $\mathbf{u}$ assuming a Gaussian radial profile controlled by three parameters, $\mathbf{v}=\mathbf{u}\,a_0 \exp\left[-(r - a_1)^2 / a_2\right]$, where $r = ||\mathbf{x}||_2$. In other words, $M$ determines the direction of the global vector field, while the Gaussian radial profile determines its magnitude. For the ion velocity, we have four configurations: (i) same velocity, i.e., $\mathbf{v}_i = \mathbf{v}$, (ii) drifted velocity, $\mathbf{v}_i = (1+\delta)\,\mathbf{v}$,  (iii) different Gaussian profile but same $M$, and  (iv) completely independent velocity (i.e., different Gaussian profile and different $M$).   Analogously, we describe the profile abundances of the two molecules with two Gaussian functions, controlled by 3 parameters  $n=\,b_0 \exp\left[-(r' - b_1)^2 / b_2\right]$, where %
$r'=\sqrt{x^2 + y^2 + \varepsilon\,z^2}$ is the radius of an ellipsoid with the prolateness/oblateness controlled by an additional parameter $\varepsilon$. We have (i) a configuration with the same density profiles for both molecules, only scaled by a relative parameterized factor, and (ii) completely independent density profiles.
Finally, the rotation of the object with respect to the observer is determined by two additional parameters, representing the intrinsic rotation angles (one of the three angles is degenerate due to the ellipsoid symmetry).
All these parameters determine the velocity vector and the density at each position $\mathbf{x}$.

To mimic the line emission of the given molecule, we use the available line velocity differences and relative intensities from $pyspeckit$. %
The lines are added using a Gaussian profile and a broadening parameter to form the spectra at the given position. The obtained intensity is scaled by the molecule abundance at the given position. The final observed spectrum is computed by summing the intensity in a uniform grid of velocity channels, an approach that mimics an optically thin radiative transfer. The result is a spectrum for each desired projected map point, i.e., a PPV cube.

The code obtained from the previous analytical model is written in JAX, a differentiable, GPU-based, and just-in-time compiled Python library. This allows us to produce a PPV cube with minimal computational impact (100-1000 models per second depending on the configurations on an NVIDIA A100 80GB), but, more importantly, to obtain the gradients of each PPV element with respect to each input parameter, thereby enabling us to optimize the parameters to minimize the difference between the observed PPV and the predicted PPV. To this aim, we use Optax \citep{deepmind2020jax} with an Adam optimizer \citep{Kingma2014}. The code is capable of optimizing both the neutral and ion PPV cubes simultaneously.

Although the model is capable of reproducing the observations with a high degree of accuracy, it is important to note that the model is an idealized representation of L1544 and, therefore, cannot reproduce some of the observed features, which are, for example, determined by the asymmetries of the actual core.
The detailed model and results are presented in \citet{Grassi2026}.

\end{appendix}

\end{document}